\documentclass[%
preprint,
amsmath,amssymb,
aps,
prfluids,
groupedaddress,
]{revtex4-2}

\usepackage{graphicx}
\usepackage{dcolumn}
\usepackage{bm}
\usepackage{lipsum} 
\usepackage[colorlinks=true,linkcolor=blue]{hyperref}%
\usepackage{color}

\begin{document}


\title{Artificial neural network approach for turbulence models: A local framework}
\author{Chenyue Xie$^{1}$}
\email{Corresponding author. Email: chenyuex@princeton.edu}
\author{Xiangming Xiong$^{1}$}
\author{Jianchun Wang$^{2}$}
\email{Email: wangjc@sustech.edu.cn}
\affiliation{\small$^{1}$Program in Applied and Computational Mathematics, Princeton University, Princeton, NJ 08544, USA\\}
\affiliation{\small$^{2}$Department of Mechanics and Aerospace Engineering, Southern University of Science and Technology, Shenzhen 518055, People's Republic of China\\}
\date{\today}
\begin{abstract}
A local artificial neural network (LANN) framework is developed for turbulence modeling. The Reynolds-averaged Navier-Stokes (RANS) unclosed terms are reconstructed by artificial neural network (ANN) based on the local coordinate system which is orthogonal to the curved walls. We verify the proposed model for the flows over periodic hills. The correlation coefficients of the RANS unclosed terms predicted by the LANN model can be made larger than 0.96 in an \emph{a priori} analysis, and the relative error of the unclosed terms can be made smaller than 18$\%$. In an \emph{a posteriori} analysis, detailed comparisons are made on the results of RANS simulations using the LANN and Spalart-Allmaras (SA) models. It is shown that the LANN model performs better than the SA model in the prediction of the average velocity, wall-shear stress and average pressure, which gives the results that are essentially indistinguishable from the direct numerical simulation (DNS) data. The LANN model trained in low Reynolds number $Re=2800$ can be directly applied in the cases of high Reynolds numbers $Re=5600$, 10595, 19000, 37000 with accurate predictions. Furthermore, the LANN model is verified for flows over periodic hills with varying slopes. These results suggest that the LANN framework has a great potential to be applied to complex turbulent flows with curved walls.


\end{abstract}

\pacs{Valid PACS appear here}
\maketitle

\section{\label{sec:level1}Introduction}
The Reynolds-averaged Navier-Stokes (RANS) simulation has been widely applied to study complex turbulent flows in industrial applications, combustion, astrophysics, and engineering problems for its low computing requirements\cite{Speziale1991,Pope2000,Durbin2018}, which can be derived by time averaging of the Navier-Stokes equations\cite{Pope2000}. Since the pioneering works of Reynolds by decomposing the instantaneous quantity into its time-averaged and fluctuating quantities\cite{Reynolds1895}, a series of model-driven approaches have been proposed to develop RANS models. These include the eddy viscosity models\cite{Boussinesq1877,Prandtl1925}, the Spalart-Allmaras model\cite{Spalart1994}, the $k-\epsilon$ model\cite{Jones1972,Launder1974,Yakhot1986,Yakhot1992}, the $k-\omega$ model\cite{Menter1994,Spalart2000,Wilcox2008}, the Reynolds stress model(RSM)\cite{Launder1975}, etc\cite{Pope2000}.

Recently, data-driven techniques have been incorporated into turbulence models \cite{Tracey2013,Duraisamy2015,Parish2016,Weatheritt2016,Ling2015,Zhang2015,Ling2016,Ling2016b,Xiao2016,Tracey2015,Brunton2016,Wangj2017,Duraisamy2017,Kutz2017,Vollant2017,Mohan2018,Raissi2018,Wu2018,Jofre2018,Ma2019,Wu2019,Zhu2019,Xie2019,Xie2019b,Xie2019c,Duraisamy2019,Fang2020,Pandey2020,Duraisamy2020,Pawar2020,Beetham2020}. The discrepancies in the Reynolds stress anisotropy tensor are reconstructed by the supervised learning\cite{Tracey2013}. Duraisamy \emph{et al.} proposed a data-driven approach to the modeling of turbulence with enforcing consistency between the data and the model\cite{Duraisamy2015,Parish2016}. Ling \emph{et al.} developed the neural network architectures embedded invariance properties in RANS simulations\cite{Ling2016}. A physics-informed Bayesian framework for quantifying and reducing model-form uncertainties in RANS simulations has been proposed by Xiao \emph{et al.}\cite{Xiao2016}. Reynolds stresses modeling discrepancies can be reconstructed by a physics-informed machine-learning approach\cite{Wangj2017}. A physics-based implicit treatment was proposed to model Reynolds stress by using machine learning techniques\cite{Wu2018}. Wu \emph{et al.} proposed a metric to quantitatively assess the conditioning of RANS equations with data-driven Reynolds stress closures\cite{Wu2019}. Furthermore, the recent progresses on data-driven turbulence models have been summarized by Duraisamy \emph{et al.}\cite{Duraisamy2019}.

In this paper, we propose a local artificial neural network (LANN) framework for reconstructing the RANS unclosed terms in the local coordinate system orthogonal to the curved wall. We find that the Reynolds unclosed terms predicted by the LANN model exhibit high accuracy in the \emph{a priori} analysis for flows over periodic hills. We also study the accuracy of the proposed LANN model in the \emph{a posteriori} tests by examining the average velocity, wall-shear stress and average pressure. These tests suggest that the LANN model is a very attractive approach for developing models of RANS unclosed terms in complex turbulent flows with curved walls.

This paper is organized as follows. Section II briefly describes the governing equations and computational method. Section III discusses the DNS database of compressible flows over periodic hills. Section IV introduces the LANN model for the reconstruction of RANS unclosed terms from the averaged flow fields. Section V presents both a \emph{priori} and \emph{a posteriori} results of the LANN model. Some discussions on the proposed LANN models are presented in Section VI. Conclusion are drawn in Section VII.

\section{Governing equations, and numerical method}
The dimensionless Navier-Stokes equations for compressible turbulence of ideal gas in the conservation form are\cite{Samtaney2001,Li2010,Wang2012,Balakumar2015,Gloerfelt2015}:

 \begin{equation}
  \frac{\partial \rho}{\partial {t}}+\frac{\partial (\rho u_{j})}{\partial x_{j}}=0,
  \label{ns1}
\end{equation}
 \begin{equation}
  \frac{\partial (\rho u_{i})}{\partial t}+\frac{\partial [\rho u_{i} u_{j}+p\delta_{ij}]}{\partial x_{j}}=\frac{1}{Re}\frac{\partial \sigma_{ij}}{\partial x_{j}},
  \label{ns2}
\end{equation}
\begin{equation}
\begin{aligned}
  \frac{\partial \mathcal{E}}{\partial {t}}+\frac{\partial [(\mathcal{E}+p)u_{j}]}{\partial x_{j}}=&\frac{1}{\alpha}\frac{\partial}{\partial{x_{j}}}(\kappa\frac{\partial T}{\partial{x_{j}}})+\frac{1}{Re}\frac{\partial (\sigma_{ij}u_{i})}{\partial x_{j}},
  \label{ns3}
\end{aligned}
\end{equation}
\begin{equation}
  p=\rho T/(\gamma Ma^{2}),
  \label{ns4}
\end{equation}
where $u_{i}$ is the $i$-th velocity component ($i=1,2,3$), $p$ is the pressure, $\rho$ is the density, and $T$ is the temperature. The viscous stress is defined by $\sigma_{ij}=2\mu S_{ij}-\frac{2}{3}\mu \delta_{ij}S_{kk}$, where $S_{ij}=\frac{1}{2}(\partial{u_{i}}/\partial{x_{j}}+\partial{u_{j}}/\partial{x_{i}})$
is the strain rate tensor. The molecular viscosity $\mu=\frac{T^{3/2(1+S)}}{T+S}$ ($S=110.4K/T_{f}$) is determined from Sutherland's law \cite{Sutherland1992}, and the thermal conductivity $\kappa$ is then calculated from the molecular viscosity with the constant Prandtl number assumption. The total energy per unit volume $\mathcal{E}$ is defined by $\mathcal{E}=\frac{p}{\gamma-1}+\frac{1}{2}\rho(u_{j}u_{j})$.

The hydrodynamic and thermodynamic variables in Eqs.~(\ref{ns1}-\ref{ns4}) are normalized by a set of reference variables: the reference velocity $U_f$, temperature $T_f$, length $L_f$, density $\rho_f$, energy per unit volume $\rho_f U_{f}^{2}$, viscosity $\mu_f$, thermal conductivity $\kappa_{f}$ and pressure $p_{f}=\rho_{f}U_{f}^{2}$. There are three reference governing parameters: the reference Reynolds number $Re\equiv \rho_{f}U_{f}L_{f}/\mu_{f}$, the reference Mach number $Ma=U_{f}/c_{f}$, and the reference Prandtl number $Pr\equiv \mu_{f}C_{p}/\kappa_{f}$. The speed of sound is defined by $c_{f}\equiv\sqrt{\gamma RT_{f}}$, where $\gamma\equiv C_{p}/C_{v}$ is the ratio of specific heat at constant pressure $C_{p}$ to that at constant volume $C_{v}$ and is assumed to be equal to 1.4. Moreover, $R\equiv C_{p}-C_{v}$ is the specific gas constant. The parameter $Pr$ is assumed to be equal to 0.7. The parameter $\alpha$ is given by $\alpha\equiv PrRe(\gamma-1)Ma^2$.

The RANS equations governing the dynamics of the mean scales, which can be obtained by projecting the physical variables into the time-averaged variables by a Reynolds operation $\bar{f}(\mathbf{x})=\lim_{T_{R} \to \infty}\frac{1}{T_{R}} \int_{t_{0}}^{t_{0}+T_{R}}f(\mathbf{x},t)dt$, where $\bar{f}$ denotes a time averaged variable, $T_{R}$ is the integration time\cite{Reynolds1895}. Favre-filtering (mass-weighted filtering: $\tilde{f}=\overline{{\rho f}}/\bar{\rho}$)\cite{Favre1965} is used to avoid additional RANS unclosed terms and simplify the treatments in the equation of conservation of mass in compressible flows. The Favre average obeys the following decomposition rules: $f=\bar{f}+f^{'}$ and $f=\tilde{f}+f^{''}$.

The dimensionless compressible governing equations for the time-averaged variables can be expressed as follows:
 \begin{widetext}
 \begin{equation}
  \frac{\partial \bar{\rho}}{\partial {t}}+\frac{\partial (\bar{\rho}\tilde{u}_{j})}{\partial x_{j}}=0,
  \label{fns1}
  \end{equation}
 \begin{equation}
  \frac{\partial (\bar{\rho}\tilde{u}_{i})}{\partial t}+\frac{\partial (\bar{\rho}\tilde{u}_{i}\tilde{u}_{j}+\bar{p}\delta_{ij})}{\partial x_{j}}-\frac{1}{Re}\frac{\partial \tilde{\sigma}_{ij}}{\partial x_{j}}=\frac{\partial\tau_{ij}}{\partial x_{j}},
  \label{fns2}
\end{equation}
\begin{equation}
\frac{\partial \tilde{\mathcal{E}}}{\partial {t}}+\frac{\partial [(\tilde{\mathcal{E}}+\bar{p})\tilde{u}_{j}]}{\partial x_{j}}-\frac{1}{Re}\frac{\partial (\tilde{\sigma}_{ij}\tilde{u}_{i})}{\partial x_{j}}-\frac{1}{\alpha}\frac{\partial}{\partial{x_{j}}}(\tilde{\kappa}\frac{\tilde{T}}{x_{j}})=\frac{\partial{C_{p}Q_{i}}}{\partial{x_{i}}}+\frac{\partial{J_{i}}}{\partial{x_{i}}},
   \label{fns3}
\end{equation}
\begin{equation}
  \bar{p}=\bar{\rho}\tilde{T}/(\gamma Ma^{2}),
  \label{fns4}
\end{equation}
  \end{widetext}
where the time-averaged total energy $\tilde{\mathcal{E}}$ is defined by $\tilde{\mathcal{E}}=\frac{\bar{p}}{\gamma-1}+\frac{1}{2}\bar{\rho}(\tilde{u}_{j}\tilde{u}_{j})$, the time-averaged viscous stress is $\tilde{\sigma}_{ij}=2\tilde{\mu} \tilde{S}_{ij}-\frac{2}{3}\tilde{\mu} \delta_{ij}\tilde{S}_{kk}$, where $\tilde{S}_{ij}=\frac{1}{2}(\partial{\tilde{u}_{i}}/\partial{x_{j}}+\partial{\tilde{u}_{j}}/\partial{x_{i}})$, and $\tilde{\mu}$ is calculated from Sutherland's law.

The RANS unclosed terms appearing on the right hand sides of Eqs.~(\ref{fns1}-\ref{fns4}) are defined as
\begin{equation}
  \tau_{ij}=-\bar{\rho}\widetilde{u^{''}_{i}u^{''}_{j}}, Q_{j}=-\bar{\rho}\widetilde{u^{''}_{j}T^{''}},
\end{equation}
where $\tau_{ij}$ is the Reynolds stress, $Q_{j}$ is the turbulent heat flux, $J_{i}$ is the triple correlation term $J_{i}=-\frac{1}{2}\bar{\rho}(\widetilde{u_{j}u_{j}u_{i}}-\widetilde{u_{j}u_{j}}\tilde{u}_{i})\approx\tau_{ij}\tilde{u}_{j}$\cite{Jiang2013,Xia2013}.

In this paper, we model the Reynolds stress $\tau_{ij}$ and turbulent heat flux $Q_{j}$, and neglect other unclosed terms. Meanwhile, we assume that the kinematic viscosity satisfies the following condition: $\overline{\sigma_{ij}}=\overline{2\rho\nu (S_{ij}-\frac{1}{3} \delta_{ij}S_{kk})}=2\bar{\rho}\nu(\widetilde{S_{ij}}-\frac{1}{3} \delta_{ij}\widetilde{S_{kk}})$, where $\nu$ is the kinematic viscosity, and the term $\frac{1}{Re}\frac{\partial \overline{\sigma_{ij}}-\widetilde{\sigma_{ij}}}{\partial x_{j}}$ would not appear in the filtered momentum equation.

\section{DNS database of compressible turbulence}
The DNS data of the compressible flows over the periodic hills (the baseline geometry of the periodic hill are depicted by piecewise cubic polynomials\cite{Mellen2000,Frohlich2005}) are obtained from the direct numerical simulation with a high-order finite difference Navier-Stokes solver ``OpenCFD-SC''\cite{Li2010,Fu2010}. A sixth-order compact finite difference scheme is used for space discretization and a third order Runge-Kutta scheme is used for time integration\cite{Lele1992,Li2010}. No-slip velocity and adiabatic conditions are imposed on the upper and lower walls for the velocity and temperature, respectively. We implement periodic boundary conditions in the streamwise $x-$ and spanwise $z-$directions. A body-fitted curvilinear gird system is used in all of the simulations\cite{Obayashi1986}. The flows are driven by a body force $F(t)$ in the streamwise direction, which is a function of time only and maintain the average mass-flux remains constant at every time step\cite{Ziefle2008,Balakumar2015}: $\frac{\partial}{\partial t}\int_{vol}\rho u dv=0$.

\begin{figure}\centering
\includegraphics[width=.8\textwidth]{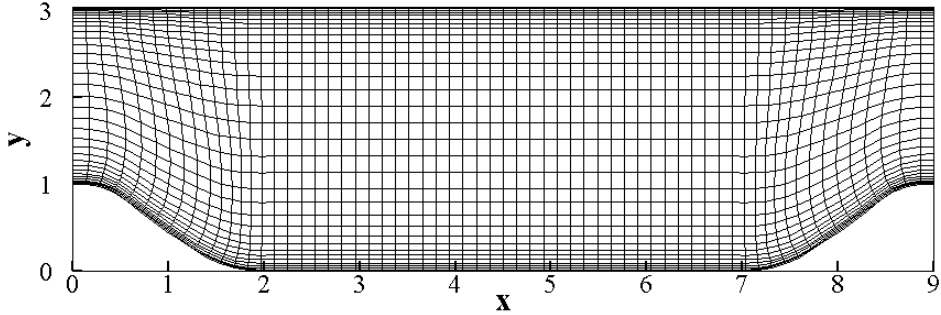}
 \caption{The configuration of the flow over periodic hill: every fourth curvilinear grid line is shown.}\label{1}
\end{figure}

The periodic-hill channel flow configuration is shown in Fig. 1\cite{Frohlich2005,Ziefle2008,Breuer2009,Balakumar2015,Gloerfelt2015,Ge2019}. The lengths are non-dimensionalized by the height of the hill $h$. The dimensionless computational domain are: $(0,L_{x})\times{0,L_{y}}\times(0,L_{z})$, where $L_{x}=9,L_{y}=3.035,L_{z}=4.5$. The cross-sectional Reynolds number over the hill crest is defined as $Re_{S}=\int_{S}(\rho u_{1})|_{x=0}dS h/(\int_{S}dS \mu_{wall})$, where $h$ is the hill height. The volumetric Reynolds number is $Re_{v}=\int_{V}\rho u_{1}dV h/(\int_{V}dV\mu_{wall})$. A relationship between $Re_{S}$ and $Re_{v}$ is $Re_{v}=0.72Re_{S}$\cite{Ziefle2008,Xia2013}. The wall temperature is fixed at $300K$. We present the DNS results at the Reynolds number $Re$ ranging from 2800 to 10595 and Mach number $Ma=0.2$ ($Ma=\int_{S}(u_{1})|_{x=0}dS/(\int_{S}dS c_{wall})$) with the grid resolution of $256\times129\times128$.

\begin{figure}\centering
\includegraphics[width=.45\textwidth]{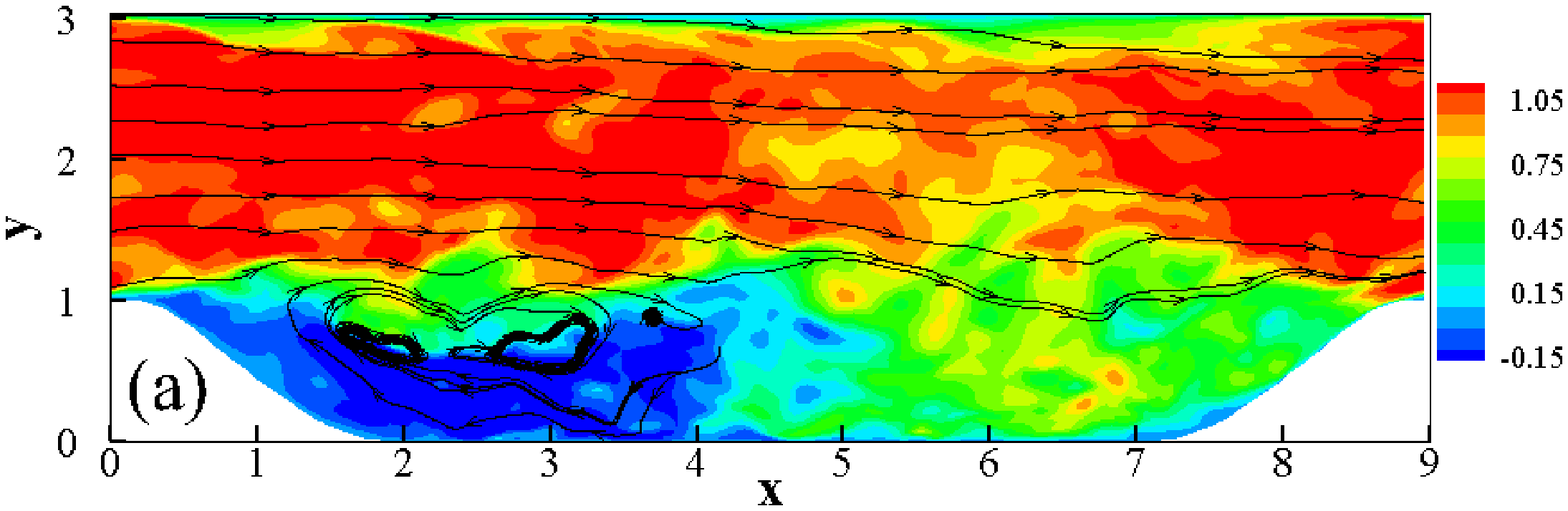}
\includegraphics[width=.45\textwidth]{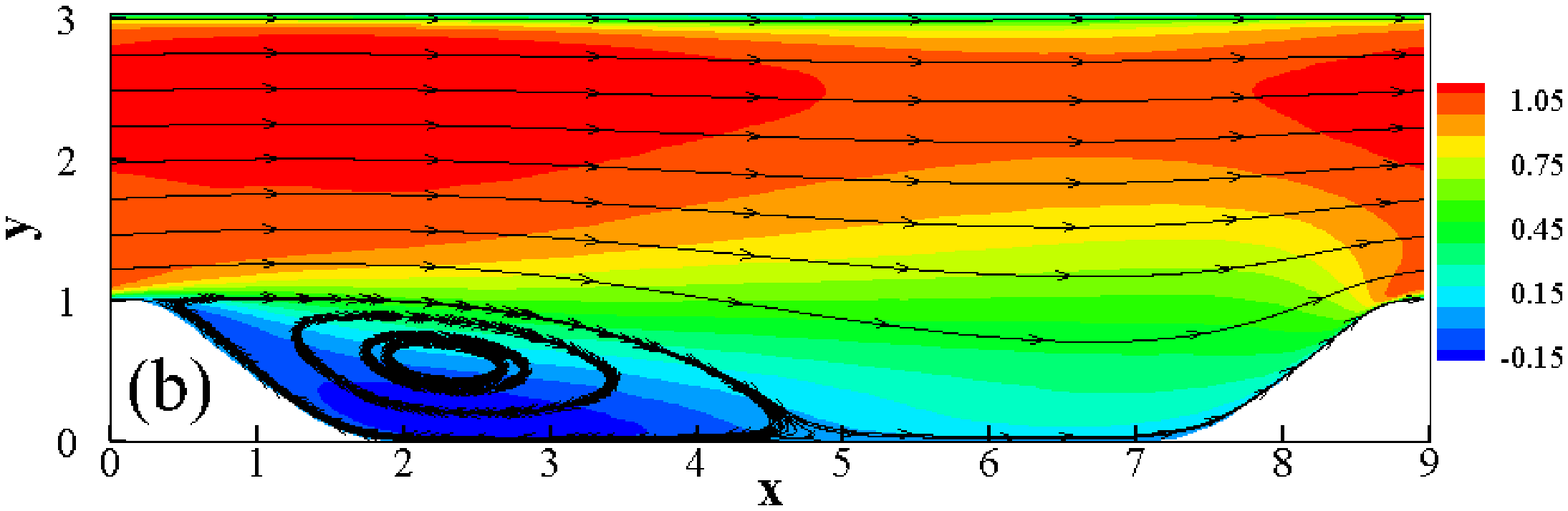}
 \caption{Contours of the streamwise velocity field at $Re=2800$: (a) the instantaneous streamwise velocity field in the (x-y) plane at $z=0$, (b) the mean streamwise velocity $\tilde{u}_{1}$ and the streamlines.}\label{2}
\end{figure}

The instantaneous streamwise velocity field at $z=0$ is shown in Fig. 2(a). The flow field is divided into two region: the reverse flow with $u_{1}<0$ and the forward flow with $u_{1}>0$. Meanwhile, the statistical average quantities are obtained by averaging the quantities in the spanwise direction and in time for 200 instantaneous flows. The contour of the mean streamwise velocity and the streamlines are displayed in Fig. 2(b). Back flow and separation occur behind the first hill. The separation and reattachment points are $x_{sep}=0.227$ and $x_{reatt}=5.34$,
respectively\cite{Ziefle2008,Breuer2009,Balakumar2015}.

\begin{figure}\centering
\includegraphics[width=.45\textwidth]{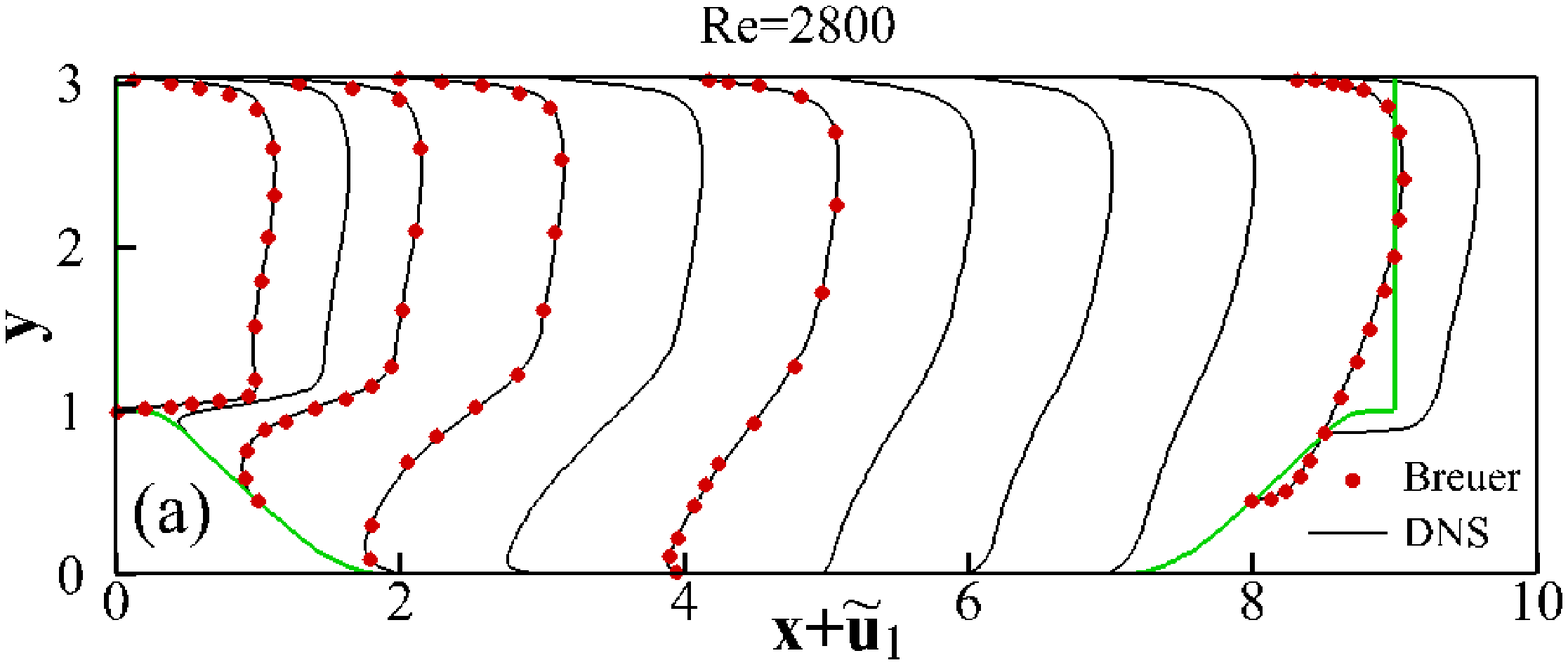}
\includegraphics[width=.45\textwidth]{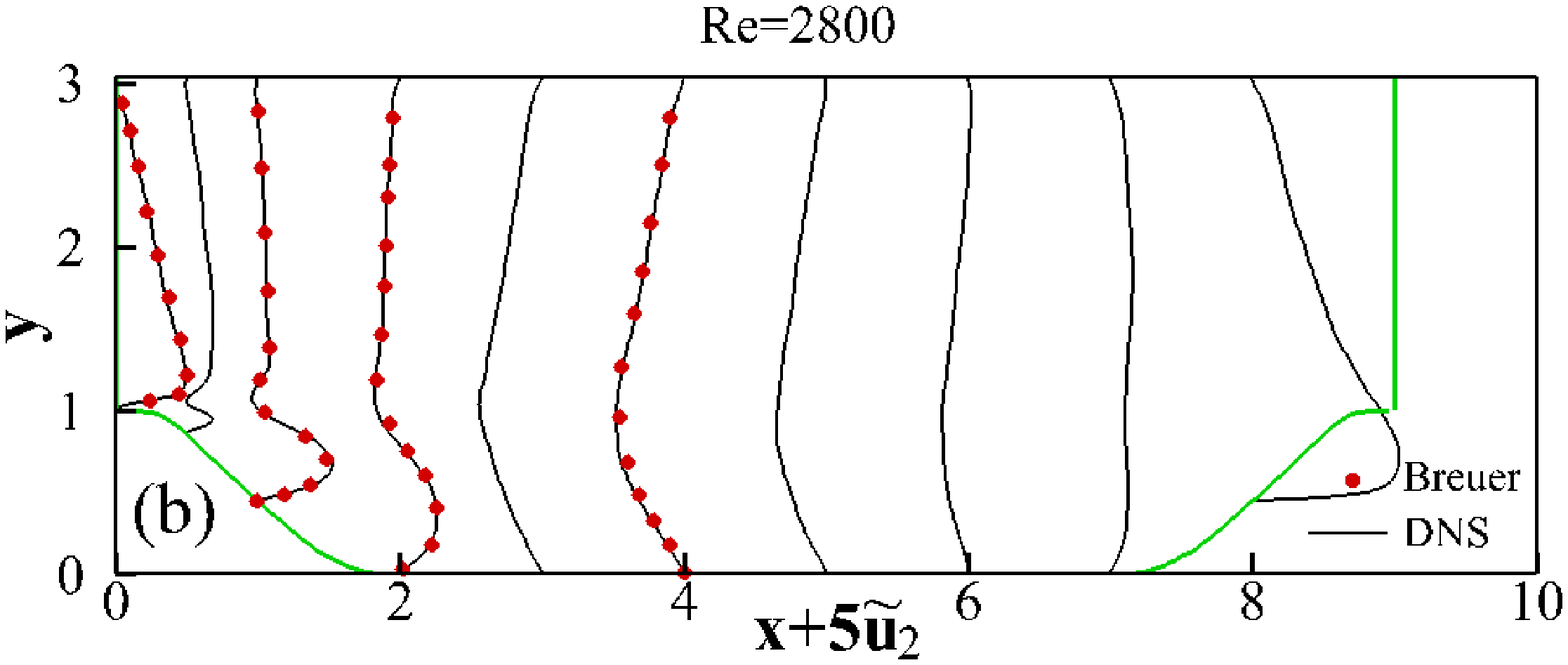}
\includegraphics[width=.45\textwidth]{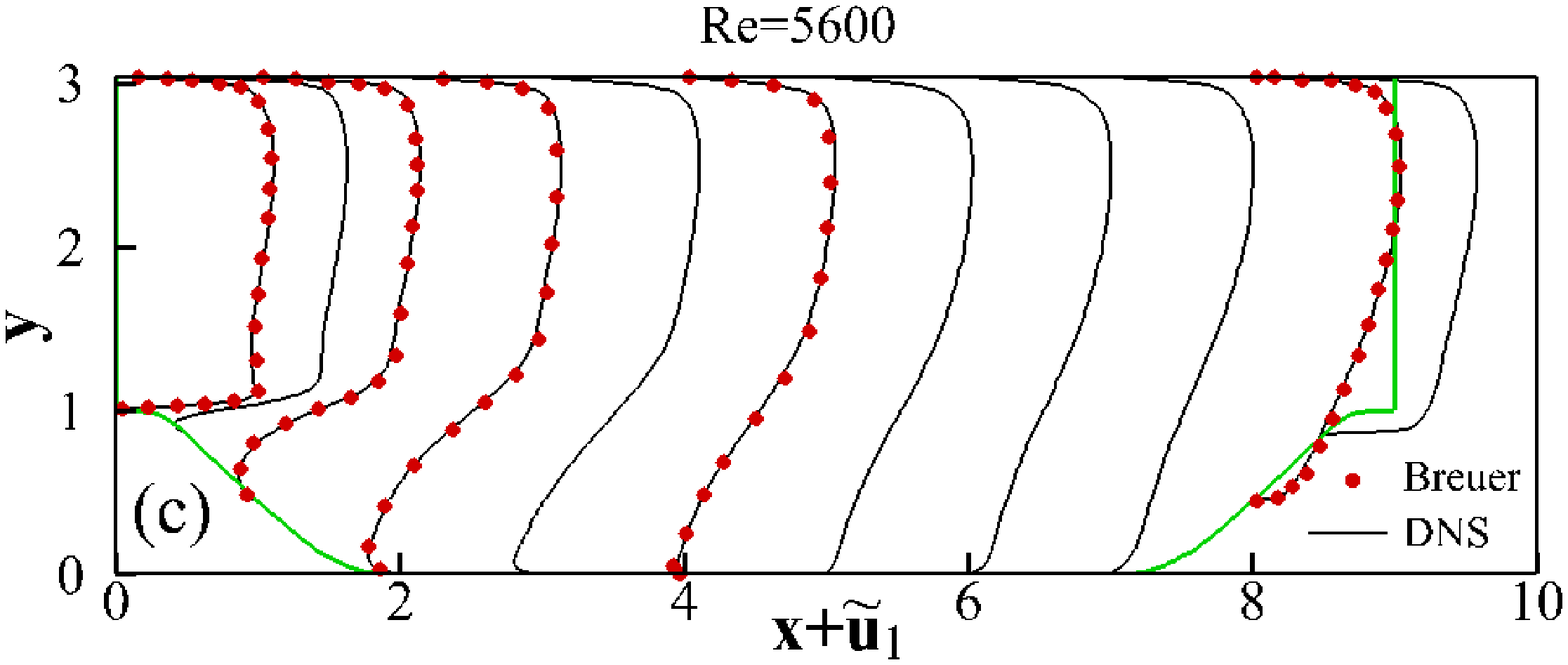}
\includegraphics[width=.45\textwidth]{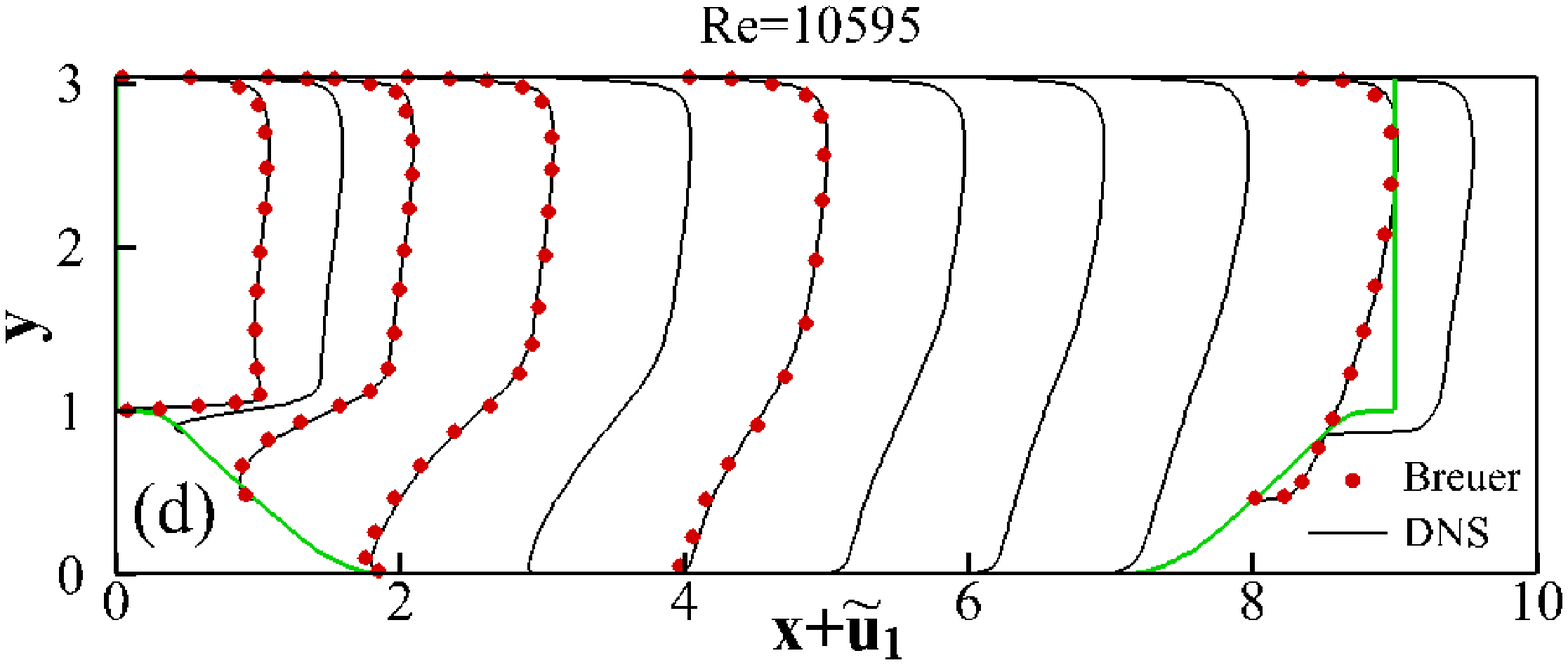}
 \caption{The mean streamwise and normal velocities $\tilde{u}_{1}$ and $\tilde{u}_{2}$ profiles at different stations $x=0$, 0.5, 1, 2, 3, 4, 5, 6, 7, 8, 8.5 with $Re=2800$, 5600, 10595: (a) $\tilde{u}_{1}$ at $Re=2800$, (b) $\tilde{u}_{2}$ at $Re=2800$, (c) $\tilde{u}_{1}$ at $Re=5600$, (d) $\tilde{u}_{1}$ at $Re=10595$.}\label{3}
\end{figure}

\begin{figure}\centering
\includegraphics[width=.8\textwidth]{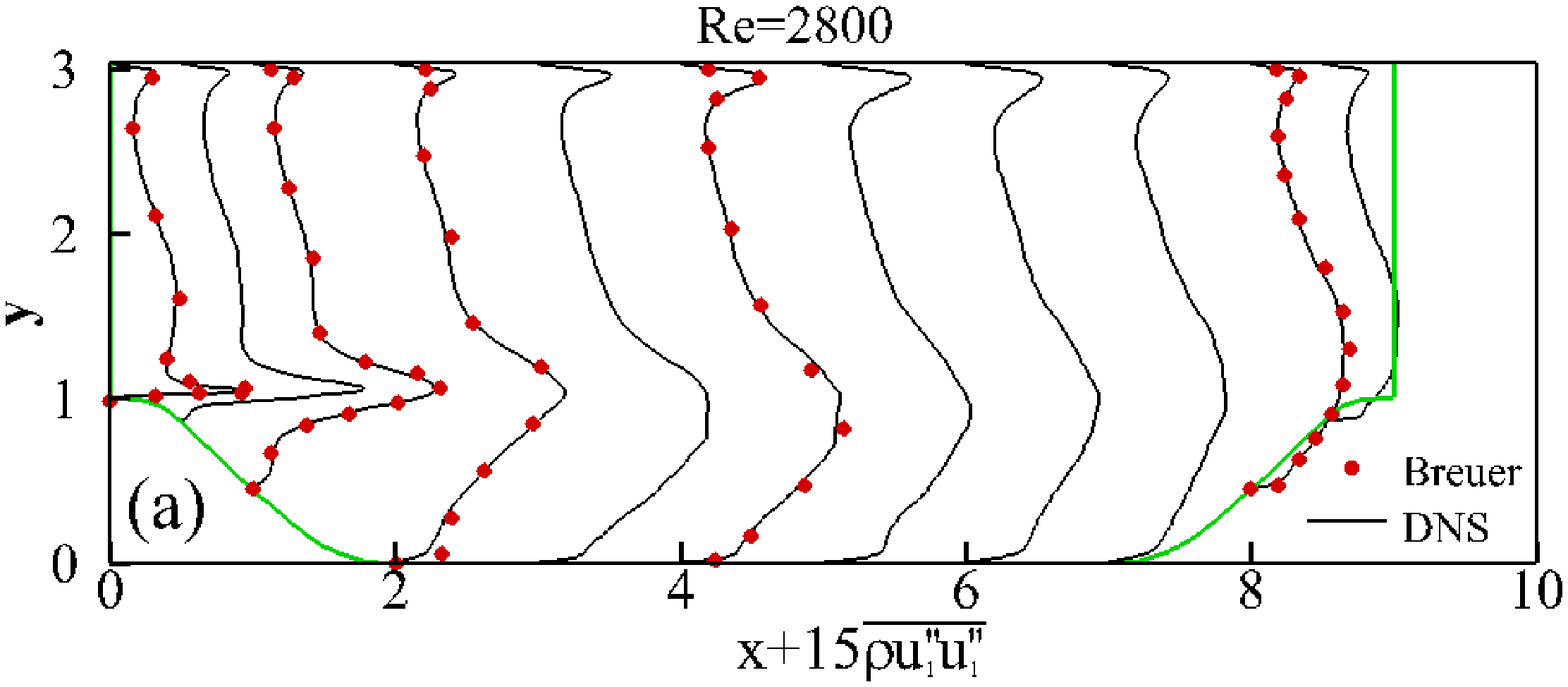}
\includegraphics[width=.8\textwidth]{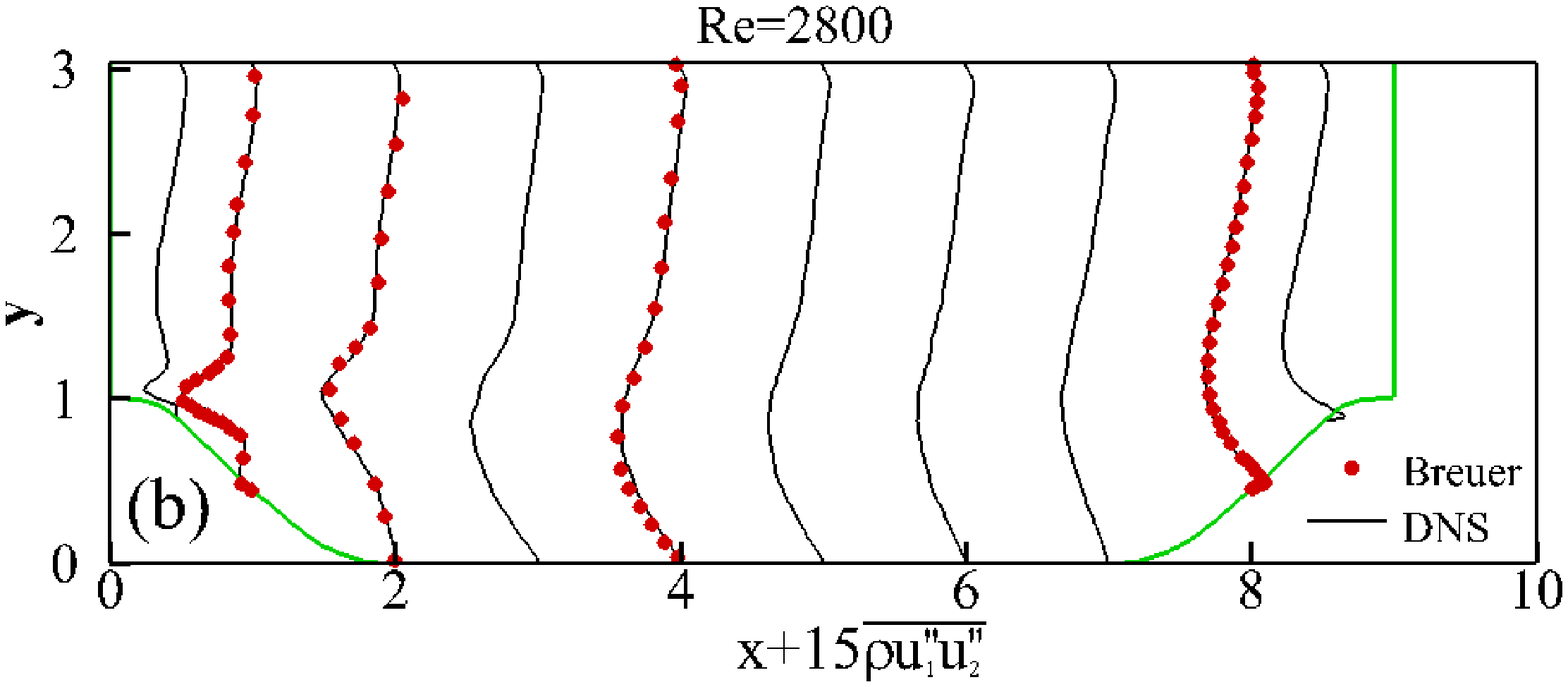}
 \caption{The Favre-averaged Reynolds stresses $\overline{\rho u_{1}^{''}u_{1}^{''}}$ and $\overline{\rho u_{1}^{'}u_{2}^{'}}$ profiles at different stations with $Re=2800$.}\label{4}
\end{figure}

\begin{figure}\centering
\includegraphics[width=.8\textwidth]{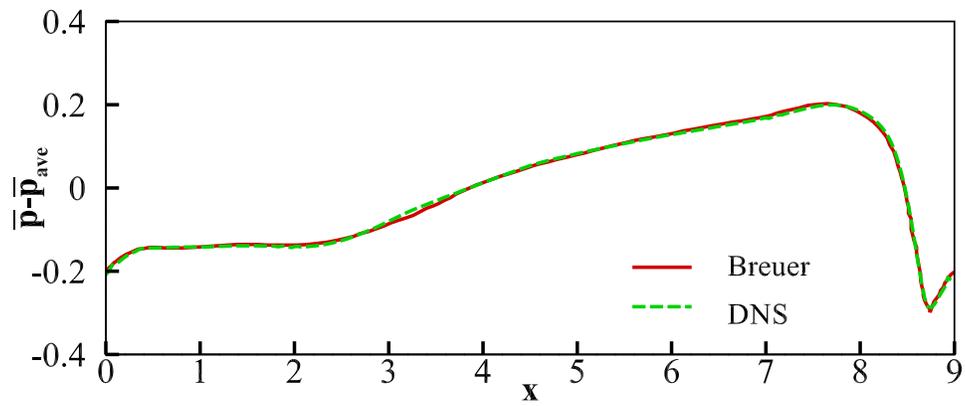}
 \caption{The averaged pressure distribution $\bar{p}-\bar{p}_{ave}$ along the lower wall for $Re=2800$.}\label{5}
\end{figure}

The comparison of the mean streamwise and normal velocities $\tilde{u}_{1}$ and $\tilde{u}_{2}$ at eleven stations $x=0$, 0.5, 1, 2, 3, 4, 5, 6, 7, 8, 8.5 for Reynolds numbers $Re=2800$, 5600, 10595 are shown in Fig.~\ref{3}. $\tilde{u}_{1}$ and $\tilde{u}_{2}$ computed from the present simulations are in excellent agreement with the DNS results of Breuer \emph{el al.}\cite{Breuer2009,Balakumar2015}. Figure 4 shows the Favre-averaged Reynolds stresses $\overline{\rho u_{1}^{''}u_{1}^{''}}$ and $\overline{\rho u_{1}^{''}u_{2}^{''}}$ at $Re=2800$. It can be seen that the solved $\overline{\rho u_{1}^{''}u_{1}^{''}}$ and $\overline{\rho u_{1}^{''}u_{2}^{''}}$ by present simulation agree well with the results of Breuer \emph{el al.}\cite{Breuer2009,Balakumar2015}. Furthermore, the averaged pressure distribution $\bar{p}-\bar{p}_{ave}$ ($\bar{p}_{ave}=\frac{1}{L_{x}}\int_{0}^{L_{x}}\bar{p}dx$) along the lower wall is shown in Fig. 5. The pressure from the present simulation is close to the previous DNS\cite{Breuer2009}. These comparisons validate the accuracy of the present direct numerical simulations.

\section{the structure of the LANN model}
\begin{figure}\centering
\includegraphics[width=.8\textwidth]{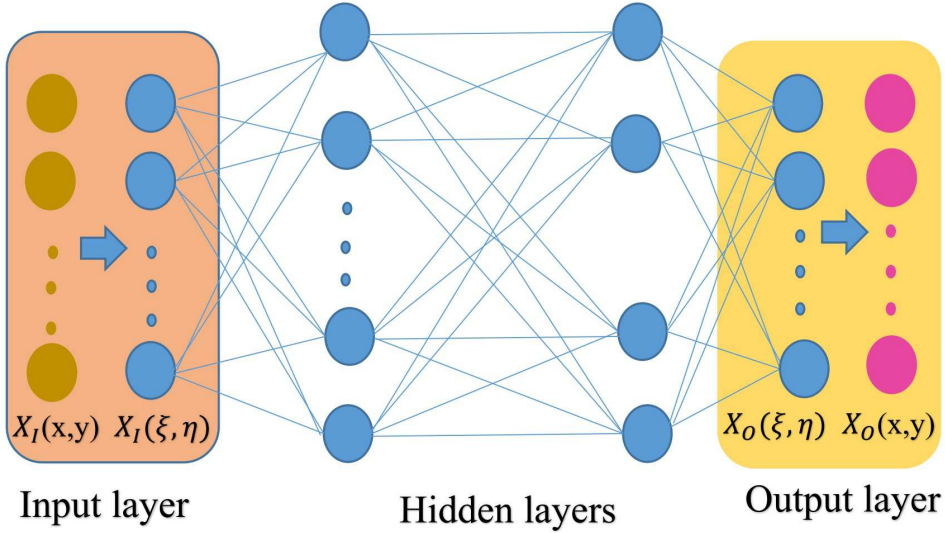}
 \caption{Schematic diagram of the ANN's network structure.}\label{6}
\end{figure}
A fully connected ANN is used to reconstruct the nonlinear relation between average input features and RANS unclosed terms $\tau_{ij}$ and $Q_{j}$. The network structure of the ANN is shown in Fig.~\ref{6}\cite{Gamahara2017,Maulik2020,Xie2020}. Neurons in layer $l$ of ANN receive signals $X^{l-1}_{j}$ from layer $l-1$ and process them with a series of linear or nonlinear mathematical operations, and then send signals $X^{l}_{i}$ to neurons in layer $l+1$\cite{Zhang1998,Demuth2014,Maulik2017}. The transfer function is calculated as
\begin{equation}
  X^{l}_{i}=\sigma(s^{l}_{i}+b^{l}_{i}),
  \label{X1}
\end{equation}
\begin{equation}
  s^{l}_{i}=\sum_{j}W^{l}_{ij}X^{l-1}_{j},
  \label{X1}
\end{equation}
where $W^{l}_{ij}$, $b^{l}_{i}$, $\sigma$ are the weight, bias parameter, and activation function, respectively. We train the ANN to update the weights and bias parameters so that the final output $X^{O}$ approximates well the RANS unclosed terms $\tau_{ij}$ and $Q_{j}$. Five ANNs are trained to predict each independent component of $\tau_{ij}$ and $Q_{j}$ separately.

In this research, the fully-connected ANN contains four layers of neurons $(M:64:32:1)$ between the set of inputs and targets. The input layer has $M$ neurons, while the output layer consists of a single neuron. Two hidden layers are activated by the Leaky-Relu activation function:
\begin{equation}
\sigma(a)=\left\{
\begin{aligned}
\quad a & , & if\ a> 0, \\
   0.2a & , & if\ a\leq0.
\end{aligned}
\right.
\end{equation}
Meanwhile, linear activation $\sigma(a)=a$ is used to the output layer. The loss function is defined by the difference between the output $X^{O}$ and the RANS unclosed terms from DNS ($\langle(X^{O}-\tau_{ij})^{2}\rangle$ or $\langle(X^{O}-Q_{j})^{2}\rangle$), where $\langle\rangle$ represents the average over the entire domain\cite{Gamahara2017}. The loss function is minimized by the back-propagation method with Adam optimizer (learning rate is 0.001)\cite{Kingma2014}.

\begin{figure}\centering
\includegraphics[width=.8\textwidth]{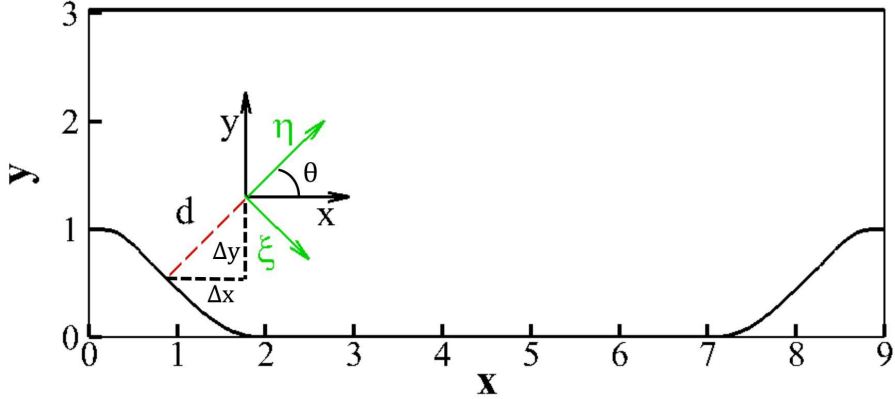}
 \caption{Transformation between the global and local reference frames in the flow over periodic hill.}\label{7}
\end{figure}

The proper choice of input variables for flows over periodic hills with varying slopes is important for the present ANN architecture to model the Reynolds stress $\tau_{ij}$ and the turbulent heat flux $Q_{j}$ accurately. As shown in the previous work\cite{Duraisamy2019}, the first-order derivatives of averaged velocities have been used to establish a functional relation between $\{\frac{\partial \tilde{u}_{i}}{\partial x_{j}},d\}$ ($d$ is the nearest distance from the walls) and the RANS unclosed terms\cite{Duraisamy2019}.

In the RANS simulations of flows over periodic hills with varying slopes, the RANS unclosed terms have been predicted by machine learning models, where the input features contain the first-order derivatives of the mean velocity and temperature in the global reference frame\cite{Duraisamy2019}. Due to that the angle between the local wall-normal direction and the global $y$ direction varies with the spatial position, it is difficult to extend the trained machine learning model to flows over periodic hills with varying slopes in the global reference frame, which makes it useless for other flows with general boundaries. In order to optimize the input features while maintaining accuracy and generality, we proposed the local artificial neural network (LANN) model, which reconstructs the nonlinear function of the input features and RANS unclosed terms $\tau_{ij}$ and $Q_{j}$ in the local coordinate system orthogonal to the nearest wall as shown in Fig.~\ref{7}. The LANN model guarantees that the nearest distance between the present point and the wall ($d$) can be measured along the $\eta$ direction of the local coordinate system, which is general for flows over periodic hills with varying slopes. A set of input variables and output variables of different ANNs are shown in Table I. As shown in Fig.~\ref{6}, the input and output features $X_{I}(x,y),X_{O}(x,y)$ in the global reference frame are transformed to $X_{I}(\xi,\eta),X_{O}(\xi,\eta)$ in local reference frame for flows over periodic hills:
\begin{equation}
\left(
\begin{array}{cc}
\frac{\partial \tilde{u}_{\xi}}{\partial \xi} & \frac{\partial \tilde{u}_{\xi}}{\partial \eta} \\
\frac{\partial \tilde{u}_{\eta}}{\partial \xi} & \frac{\partial \tilde{u}_{\eta}}{\partial \eta} \\
\end{array}
\right)=A\left(
\begin{array}{cc}
\frac{\partial \tilde{u}_{x}}{\partial x} & \frac{\partial \tilde{u}_{x}}{\partial y} \\
\frac{\partial \tilde{u}_{y}}{\partial x} & \frac{\partial \tilde{u}_{y}}{\partial y} \\
\end{array}
\right)A^{T}
  \label{trans1}
\end{equation}
\begin{equation}
\left(
  \begin{array}{cc}
   \frac{\partial \tilde{T}}{\partial \xi}  & \frac{\partial \tilde{T}}{\partial \eta} \\
  \end{array}
\right)
=\left(
  \begin{array}{cc}
   \frac{\partial \tilde{T}}{\partial x}  & \frac{\partial \tilde{T}}{\partial y} \\
  \end{array}
\right)A^{T}
  \label{trans2}
\end{equation}
\begin{equation}
\left(
\begin{array}{cc}
\tau_{\xi\xi}  & \tau_{\xi\eta} \\
\tau_{\eta\xi} & \tau_{\eta\eta} \\
\end{array}
\right)=A\left(
\begin{array}{cc}
\tau_{xx}  & \tau_{xy} \\
\tau_{yx}  & \tau_{yy} \\
\end{array}
\right)A^{T}
  \label{trans3}
\end{equation}
\begin{equation}
\left(
  \begin{array}{cc}
   Q_{\xi T}  & Q_{\eta T} \\
  \end{array}
\right)
=\left(
  \begin{array}{cc}
   Q_{x T}  & Q_{y T} \\
  \end{array}
\right)A^{T}
  \label{trans4}
\end{equation}
where $A=\left(\begin{array}{cc}
sin(\theta) & -cos(\theta) \\
cos(\theta) & sin(\theta) \\
\end{array}
\right), $ $cos(\theta)=\frac{\Delta x}{r}$, $sin(\theta)=\frac{\Delta y}{r}$, $r=\sqrt{\Delta x^{2}+\Delta y^2}$.

\begin{table}[tbp]
\centering
\begin{tabular}{lcc}
\hline\hline
ANN &Inputs  &Outputs\\ \hline
ANN1 &$\frac{\partial \tilde{u}_{\xi}}{\partial \xi},\frac{\partial \tilde{u}_{\xi}}{\partial \eta},\frac{\partial \tilde{u}_{\eta}}{\partial \xi},\frac{\partial \tilde{u}_{\eta}}{\partial \eta},\bar{\rho},d,\bar{\mu}$    &$\tau_{\xi\eta}$\\ \hline
ANN2 &$\frac{\partial \tilde{u}_{\xi}}{\partial \xi},\frac{\partial \tilde{u}_{\xi}}{\partial \eta},\frac{\partial \tilde{u}_{\eta}}{\partial \xi},\frac{\partial \tilde{u}_{\eta}}{\partial \eta},\frac{\partial \tilde{T}}{\partial \xi},\frac{\partial \tilde{T}}{\partial \eta},\bar{\rho},d,\bar{\mu}$    &$Q_{\xi}$\\ \hline\hline
\end{tabular}
\caption{Set of inputs and outputs for the ANNs.}
\end{table}

In order to increase the robustness of the ANN training, the first-order derivatives of the mean velocity and temperature in $X_{I}$ are normalized by their root mean square (rms) values, which is similar to the previous data-driven strategies\cite{Ling2015,Xiao2016,Wangj2017,Vollant2017,Maulik2017,Xie2020}:
\begin{eqnarray}
&& Z_{I}=X_{I}/X_{I}^{rms}.
\label{Xi}
\end{eqnarray}

\begin{figure}\centering
\includegraphics[width=.8\textwidth]{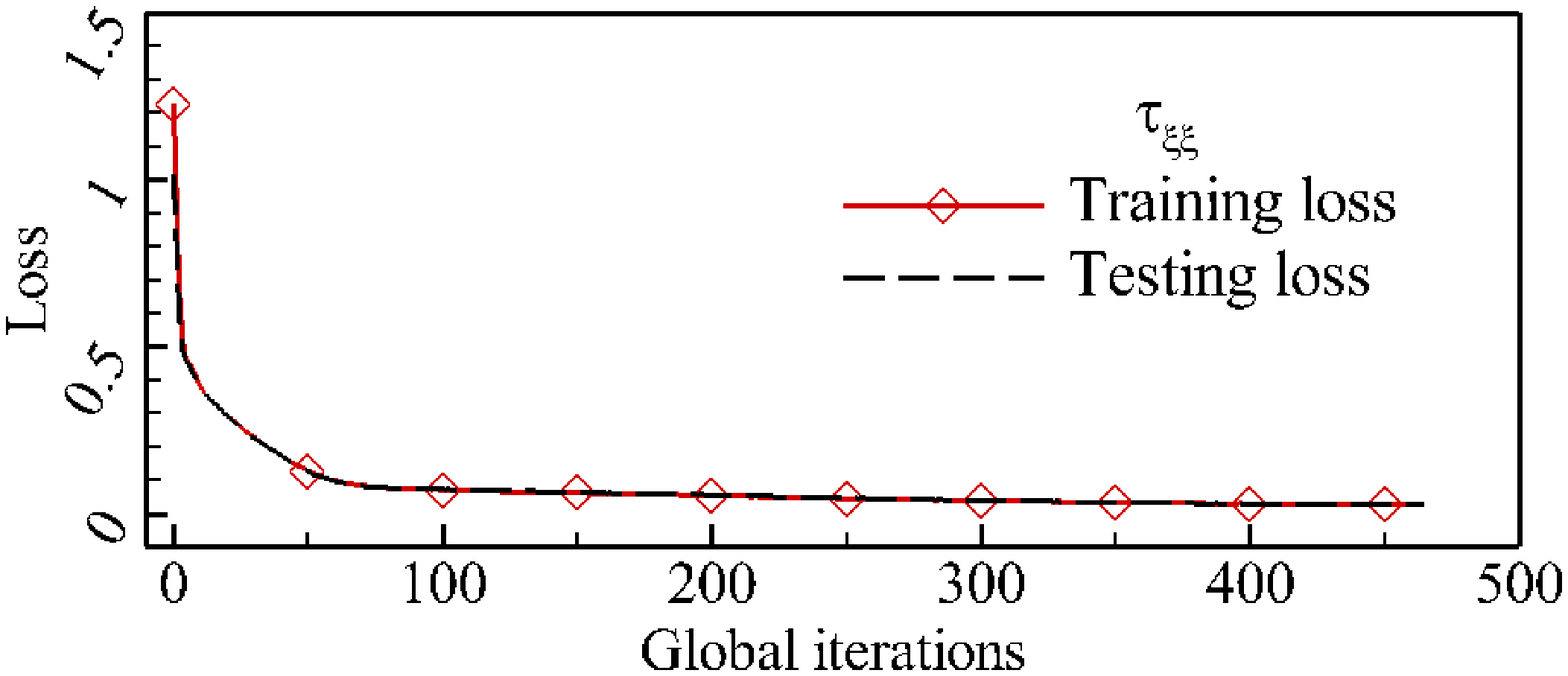}
 \caption{Learning curves of the proposed LANN model of the unclosed Reynolds stress $\tau_{\xi\xi}$.}\label{8}
\end{figure}

Besides, we suppress overfitting with a cross-validation. The performance of the model is estimated by the data which have not been used for training. In this research, the inputs and outputs of the LANN model are the mean flow features and the RANS unclosed terms $\tau_{ij},Q_{j}$, respectively, which are obtained from the DNS data. The three-dimensional (3D) DNS data are generated using $256\times129\times128$ degrees of freedom while the two-dimensional (2D) RANS is performed at the grid resolutions of $256\times129$ and $128\times65$. Finally, the network is trained by the Adam algorithm\cite{Kingma2014} with early stopping (if validation errors did not decrease improve for 10 epochs, the training would exit with the best model corresponding to the lowest validation loss until then)\cite{Maulik2020}. The learning rate and batch size of the ANN are 0.001 and 1000, respectively. The total data for ANN training are $32639$ grid points at $Re=2800$. 70$\%$ of data is for training, and the left 30$\%$ is for testing. The training and testing losses show similar behavior and correlate closely after 100 global iterations as shown in Fig.~\ref{8}, which implies that the learning process is reasonable.

\section{test results of the LANN model}

In this section, we conduct both \emph{a priori} and \emph{a posteriori} tests to evaluate the performance of the LANN model for flows over periodic hills. The LANN model trained at $Re=2800$ is used to produce reliable and repeatable predictions at $Re=2800$, 5600, 10595, 19000, 37000. We calculated the correlation coefficients and relative errors of the predicted RANS unclosed terms $\tau_{ij}$ and $Q_{j}$ in the \emph{a priori} test. In the \emph{a posteriori} test, results of the RANS simulations with the LANN model are compared with the SA model and the DNS database. It is shown that the RANS simulations with the proposed LANN model can predict the statistics of the averaged DNS data with high accuracy.

\subsection{\emph{A priori} tests}

We evaluate the performance of the LANN model by calculating the correlation coefficient $C(R)$ and the relative error $E_{r}(R)$ of $\tau_{ij}$ and $Q_{j}$. $C(R)$ and $E_{r}(R)$ are defined, respectively, by
\begin{equation}
C(R)=\frac{\langle(R-\langle R\rangle)(R^{model}-\langle R^{model}\rangle)\rangle}{(\langle(R-\langle R\rangle)^2\rangle\langle(R^{model}-\langle R^{model}\rangle)^2\rangle)^{1/2}},
   \label{corr}
\end{equation}
\begin{equation}
E_{r}(R)=\frac{\sqrt{\langle(R-R^{model})^2\rangle}}{\sqrt{\langle R^2\rangle}},
   \label{rela}
\end{equation}
where $\langle\cdot\rangle$ denotes averaging over the volume.

\begin{table}[tbp]
\setlength{\tabcolsep}{0.8mm}{
\begin{tabular}{lccccc}
\hline\hline
$C$  &$\tau_{\xi\xi}$ &$\tau_{\xi\eta}$ &$\tau_{\eta\eta}$ &$Q_{\xi}$ &$Q_{\eta}$\\ \hline
train &0.967 &0.992 &0.993 &0.980 &0.986\\
test  &0.966 &0.992 &0.993 &0.979 &0.987\\ \hline\hline
\end{tabular}}
\setlength{\tabcolsep}{0.8mm}{
\begin{tabular}{lccccc}
\hline\hline
$E_{r}$  &$\tau_{\xi\xi}$ &$\tau_{\xi\eta}$ &$\tau_{\eta\eta}$ &$Q_{\xi T}$ &$Q_{\eta T}$\\ \hline
train &0.140 &0.111 &0.082 &0.177 &0.164\\
test  &0.141 &0.110 &0.084 &0.181 &0.162\\ \hline\hline
\end{tabular}}
\caption{Correlation coefficient (C) and relative error ($E_{r}$) of $\tau_{\xi\xi}$, $\tau_{\xi\eta}$, $\tau_{\eta\eta}$, $Q_{\xi T}$, and $Q_{\eta T}$ for the LANN model in the local reference frame at $Re=2800$.}
\end{table}
Table~II shows the correlation coefficients and relative errors of $\tau_{\xi\eta}$ and $Q_{\xi}$ in the local reference frame for LANN model in both training and testing sets at $Re=2800$. The difference between the results of training and testing sets is small, which implies that the training process of ANN is not overfitting. The correlation coefficients are larger than 0.96, and the relative errors are less than 0.18 for the LANN model.

\subsection{\emph{A posteriori} tests}

We evaluate the performance of the LANN model for flows over periodic hills with varying slopes at $Re=2800$, 5600, 10595,19000, 37000. Furthermore, in order to show that the LANN model can be applied to flows over periodic hills with varying slopes, the \emph{a posteriori} studies of the LANN model applied to flow over periodic hill with the total horizontal length of the domain $L_{x}=3.858\alpha+5.142$, where $\alpha=1.5$ is conducted\cite{Xiao2020}. Here, $\alpha$ controls the width of the hill, and the length of the flat section between the hills is 5.142, which is kept constant. The two dimensional Reynolds-averaged Navier-Stokes equations are solved with a finite volume solver ``OpenCFD-EC'' developed by Li $\emph{et al.}$\cite{Gao2005,Li2005}. The spatial gradients are calculated with a second-order accurate discretization. The temporal advancement of the equations is achieved using an implicit LU-SGS method. The flow is set to be periodic in the streamwise direction. No-slip condition and adiabatic condition are set at walls for the velocity and temperature, respectively.

Eddy viscosity turbulence models have been widely used for aeronautical, meteorological, and other applications\cite{Pope2000}. The Boussinesq hypothesis is applied to establish the relation between RANS unclosed terms and the first-order derivatives of mean velocity and temperature\cite{Boussinesq1877}. The traceless part of the Reynolds stress $\tau_{ij}$ is proportional to the product of the mean strain rate tensor $\tilde{S}_{ij}$ and the eddy viscosity $\mu_{t}$. The Reynolds stress $\tau_{ij}$ and turbulent heat flux $Q_{j}$ are\cite{Spalart1994}:
$\tau_{ij}=-\mu_{t}(2\tilde{S}_{ij}-\frac{2}{3}\frac{\partial \tilde{u}_{k}}{\partial x_{k}})+\frac{2}{3}\bar{\rho} k\delta_{ij}$ (where the last term is generally ignored for one-equation models because $k$ is not readily available), $Q_{j}=-\frac{C_{p}\mu_{t}}{Pr_{T}}\frac{\partial \tilde{T}}{\partial x_{j}}$. The eddy viscosity is given by
\begin{equation}
 \mu_{t}=\rho\nu_{t}f_{\nu 1}.
  \label{SA}
\end{equation} The Spalart-Allmaras (SA) model solves a transport equation for $\nu_{t}$. The governing equation for the intermediate variable $\nu_{t}$ is\cite{Spalart1994}:
\begin{equation}
 \frac{\partial \nu_{t}}{\partial t}+\frac{\partial}{\partial x_{j}}(\nu_{t}\tilde{u}_{j})=c_{b1}S_{t}\nu_{t}+\frac{1}{\sigma}[\frac{\partial}{\partial x_{j}}((\tilde{\nu}+\nu_{t})\frac{\partial \nu_{t}}{\partial x_{j}})+c_{b2}\frac{\partial \nu_{t}}{\partial x_{j}}\frac{\partial \nu_{t}}{\partial x_{j}}]-c_{w1}f_{w}(\frac{\nu_{t}}{d})^2,
  \label{SA}
\end{equation}
where $S_{t}=\Omega+\frac{\nu_{t}}{k^2 d^2}f_{\nu 2}$, $f_{\nu 1}=\frac{\chi^{3}}{\chi^{3}+c^{3}_{\nu 1}}$, $f_{\nu 2}=1-\frac{\chi}{1+\chi/f_{\nu 1}}$, $\chi=\frac{\rho \nu_{t}}{\mu}$, $f_{w}=g[\frac{1+C^{6}_{w3}}{g^{6}+C^{6}_{w3}}]^{1/6}$, $g=r+C_{w2}(r^{6}-r)$, $r=\frac{\nu_{t}}{S_{t}k^{2}d^{2}}$, $\Omega$ is the magnitude of the vorticity vector, and $d$ is the nearest distance from the walls. Meanwhile, the model coefficients in the SA model are $\sigma=2/3$, $C_{b1}=0.1355$, $C_{b2}=0.622$, $\kappa=0.41$, $C_{w1}=\frac{c_{b1}}{\kappa^{2}}+(1+C_{b2})/\sigma$, $C_{w2}=0.3$, $C_{w3}=2$, $C_{\nu 1}=7.1$, and $C_{w1}=C_{b1}/k^{2}+(1+C_{b2})/\sigma$. The periodic boundary condition is applied in the streamwise x-direction for $\nu_{t}$. $\nu^{wall}_{t}=0$ is imposed on the upper and lower wall boundaries.

\begin{figure}\centering
\includegraphics[width=.45\textwidth]{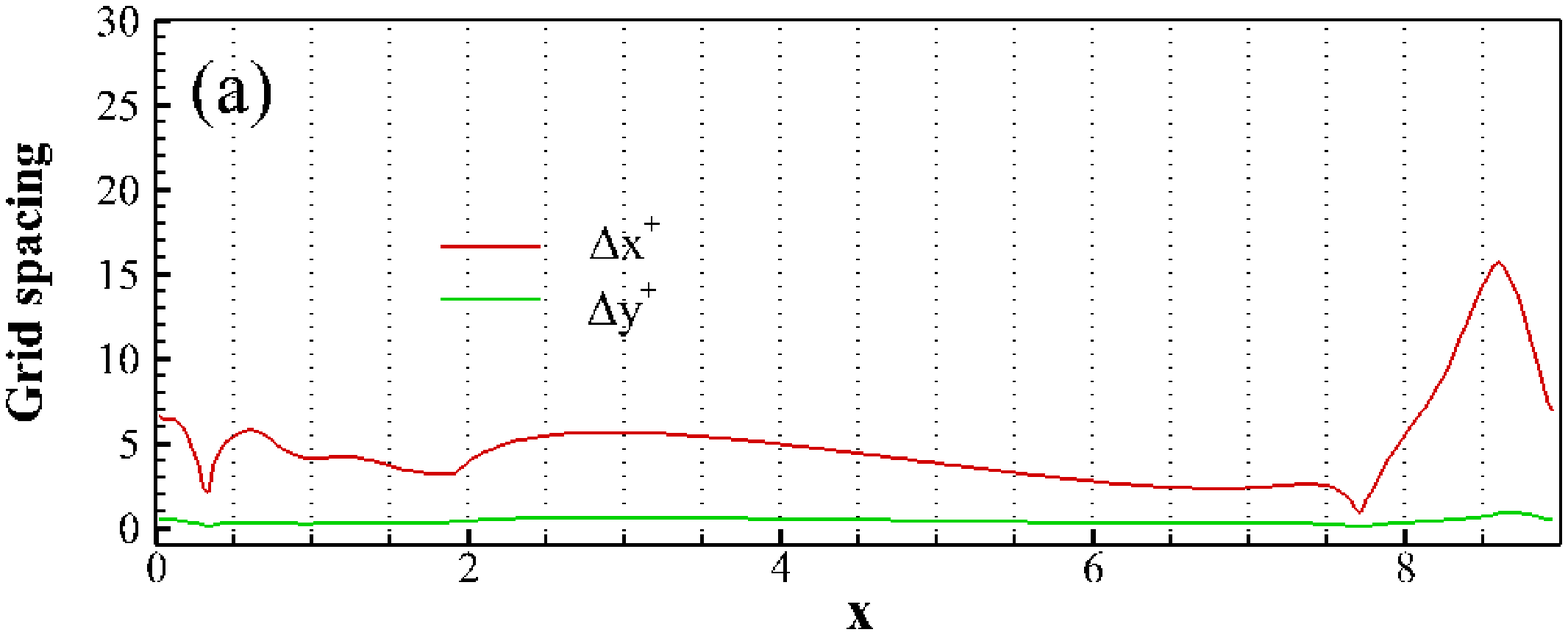}
\includegraphics[width=.45\textwidth]{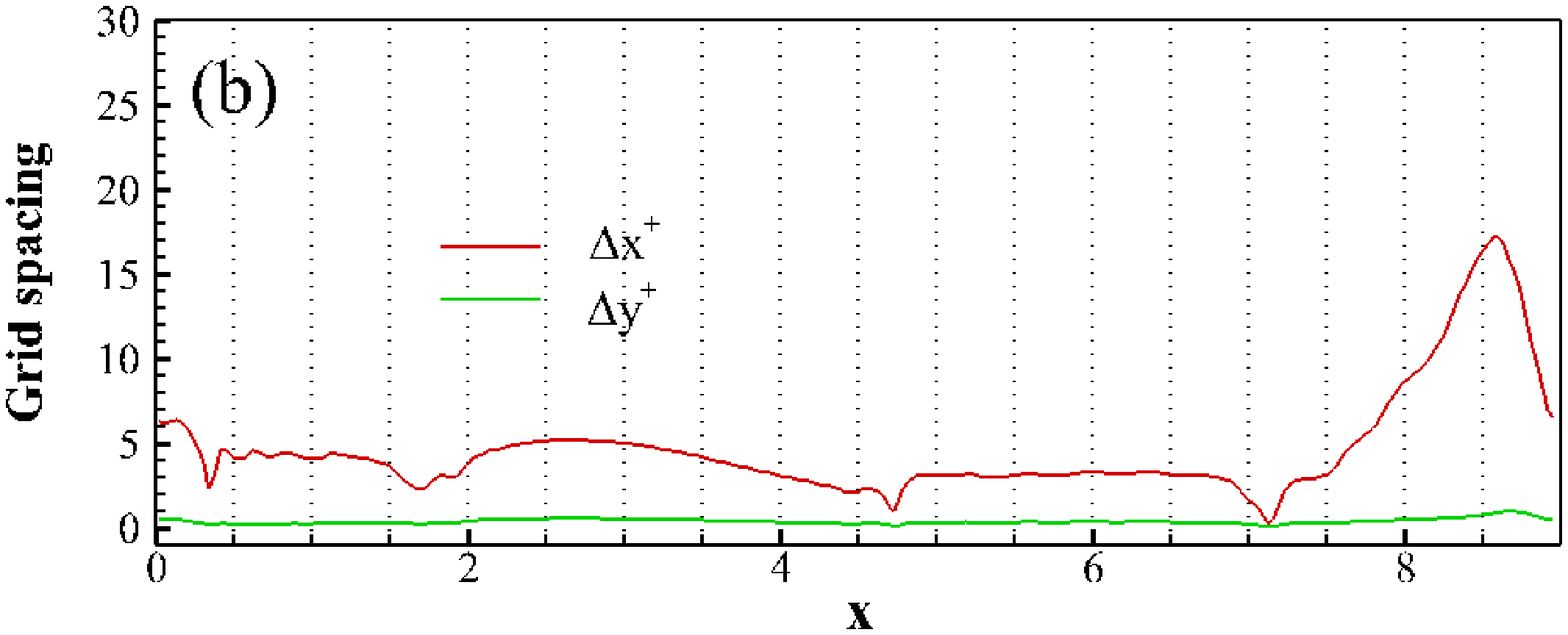}
 \caption{Grid spacing in wall units along the bottom wall of RANS simulations with the SA and LANN models at $Re=2800$ with a grid resolution of $256\times129$ and $L_{x}=9.0$: (a) SA model, (b) LANN model.}\label{9}
\end{figure}

\begin{figure}\centering
\includegraphics[width=.45\textwidth]{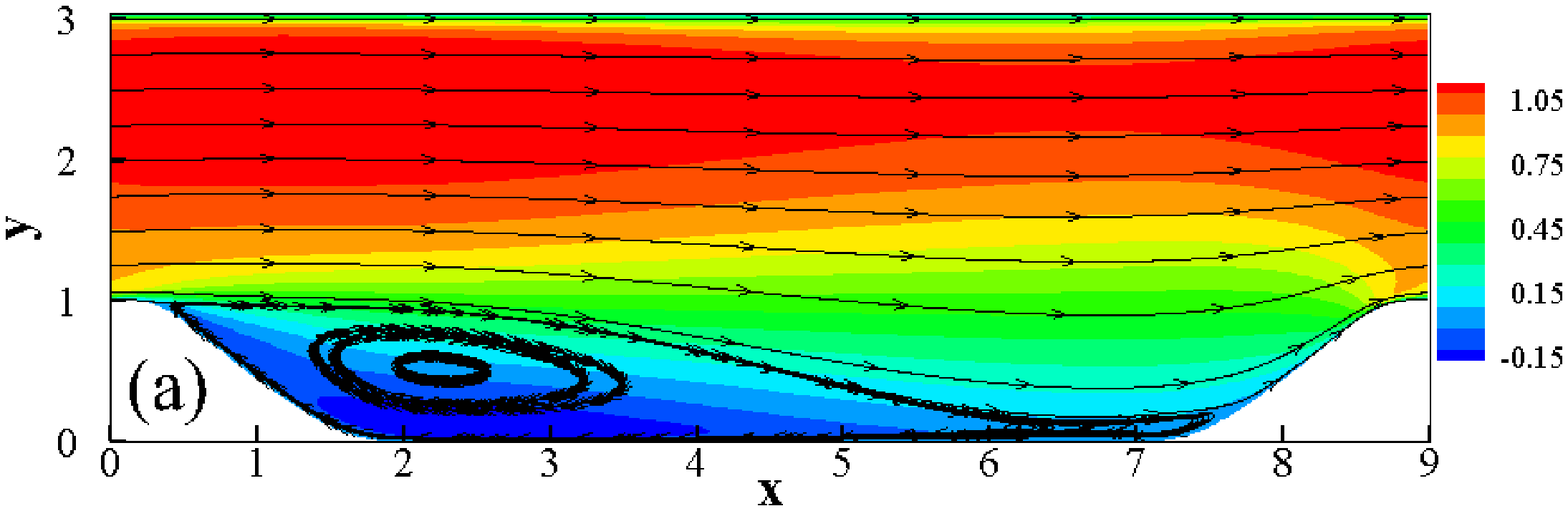}
\includegraphics[width=.45\textwidth]{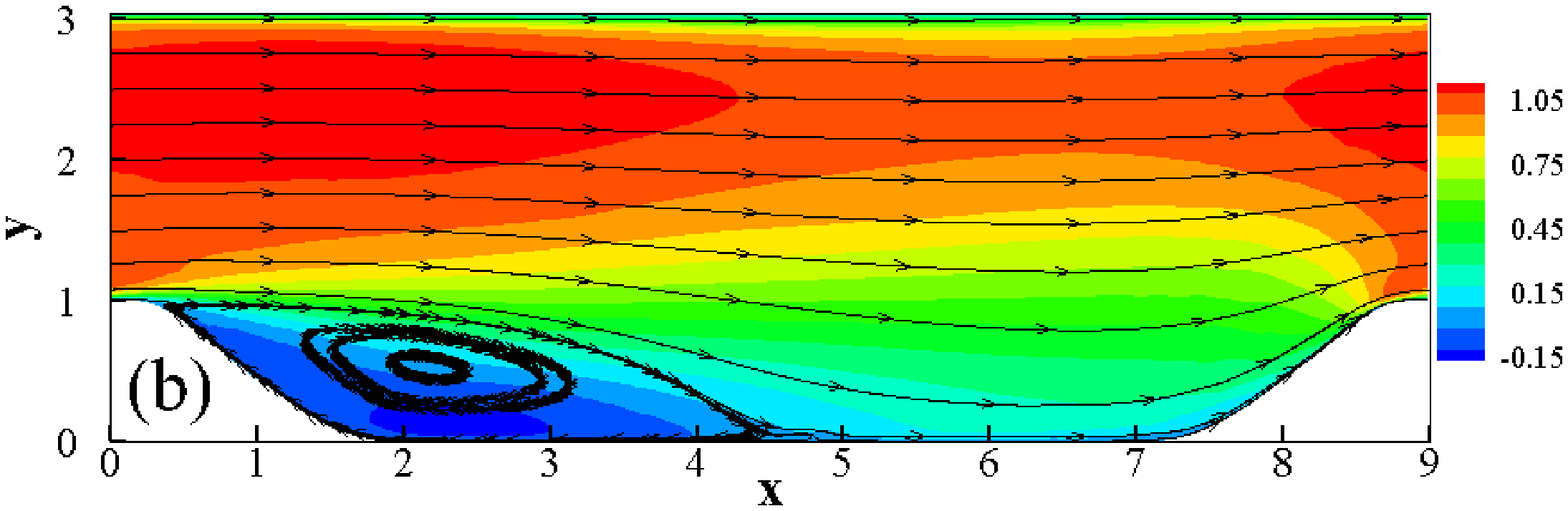}
 \caption{Hill flow contours of the mean streamwise velocity $\tilde{u}$ and the streamlines at $Re=2800$ with a grid resolution of $256\times129$ and $L_{x}=9.0$: (a) SA model, (b) LANN model.}\label{10}
\end{figure}

The performances of the LANN model are evaluated by calculating the average velocity, wall-shear stress and average pressure. In the \emph{a posteriori} tests, the initial conditions for the LANN model are generated from the steady-state flow fields calculated by the RANS simulation with the SA model. The variations of the grid spacing in wall units in the $x$ and $y$ directions along the bottom wall for RANS simulations with the SA and LANN models are shown in Fig.~\ref{9}. The maximum grid spacings in the $x$ and $y$ directions are located at the downstream wall due to a large increase in the friction velocity in this region. Figure~\ref{10} shows the mean streamwise velocity contours and the streamlines from RANS simulations with the SA and LANN models. The separation and reattachment points are at $x_{sep}=0.316,0.346$ and $x_{reatt}=7.70,4.80$ for the SA and LANN models, respectively. The separation point at $x_{sep}=0.316$ predicted by the SA model is slightly closer to the DNS result ($x_{sep}=0.227$). The reattachment point at $x_{reatt}=4.80$ predicted by the LANN model is much closer to the DNS result ($x_{reatt}=5.34$).

\begin{figure}\centering
\includegraphics[width=.8\textwidth]{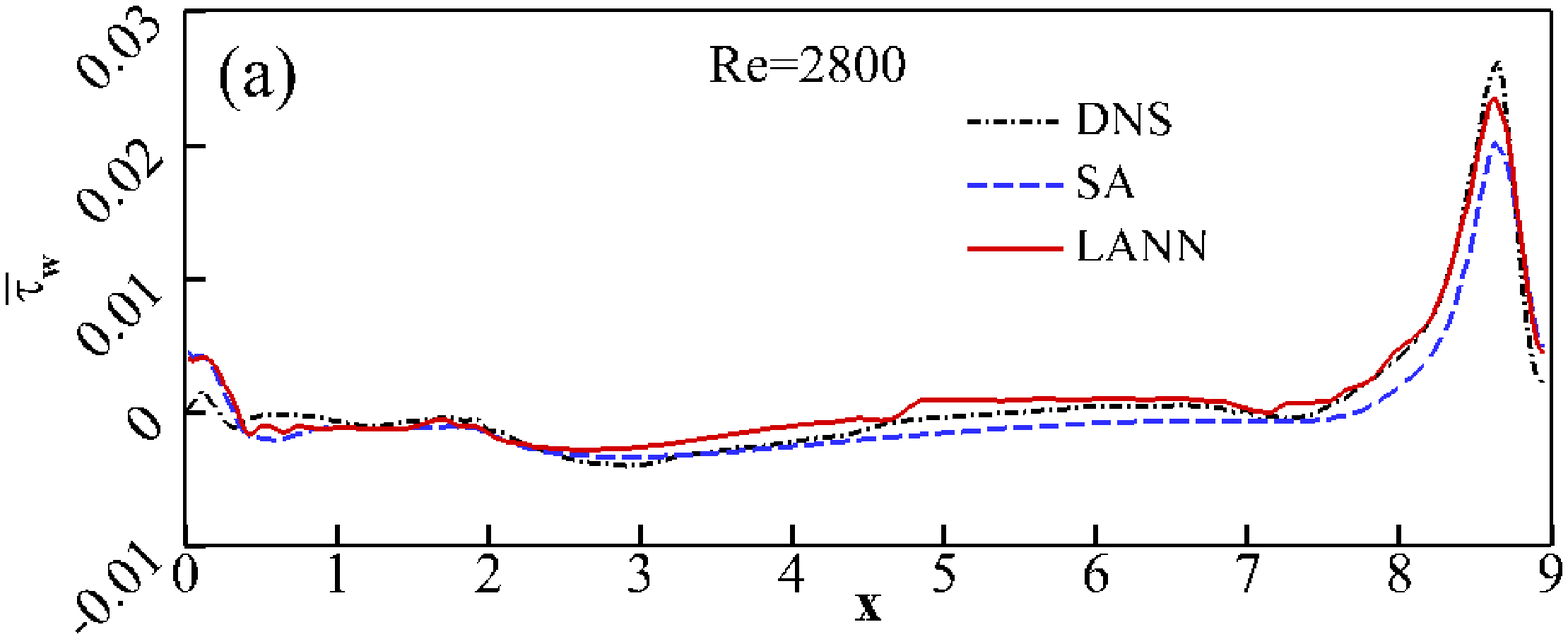}
\includegraphics[width=.8\textwidth]{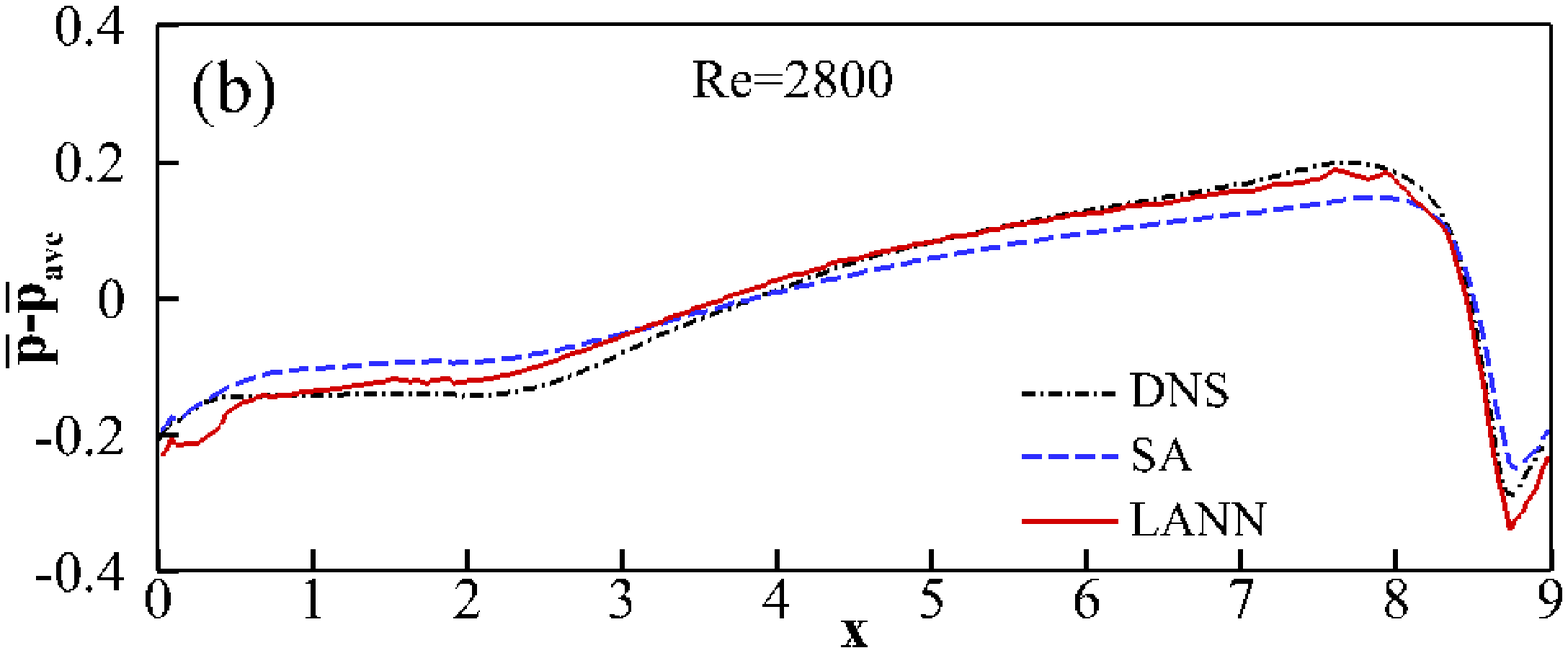}
 \caption{Profiles of the averaged wall shear stress $\tilde{\tau}_{w}$ and the averaged pressure distribution $\bar{p}$ at $Re=2800$ with a grid resolution of $256\times129$ and $L_{x}=9.0$: (a) $\tilde{\tau}_{w}$, (b) $\bar{p}$.}\label{11}
\end{figure}

\begin{figure}\centering
\includegraphics[width=.8\textwidth]{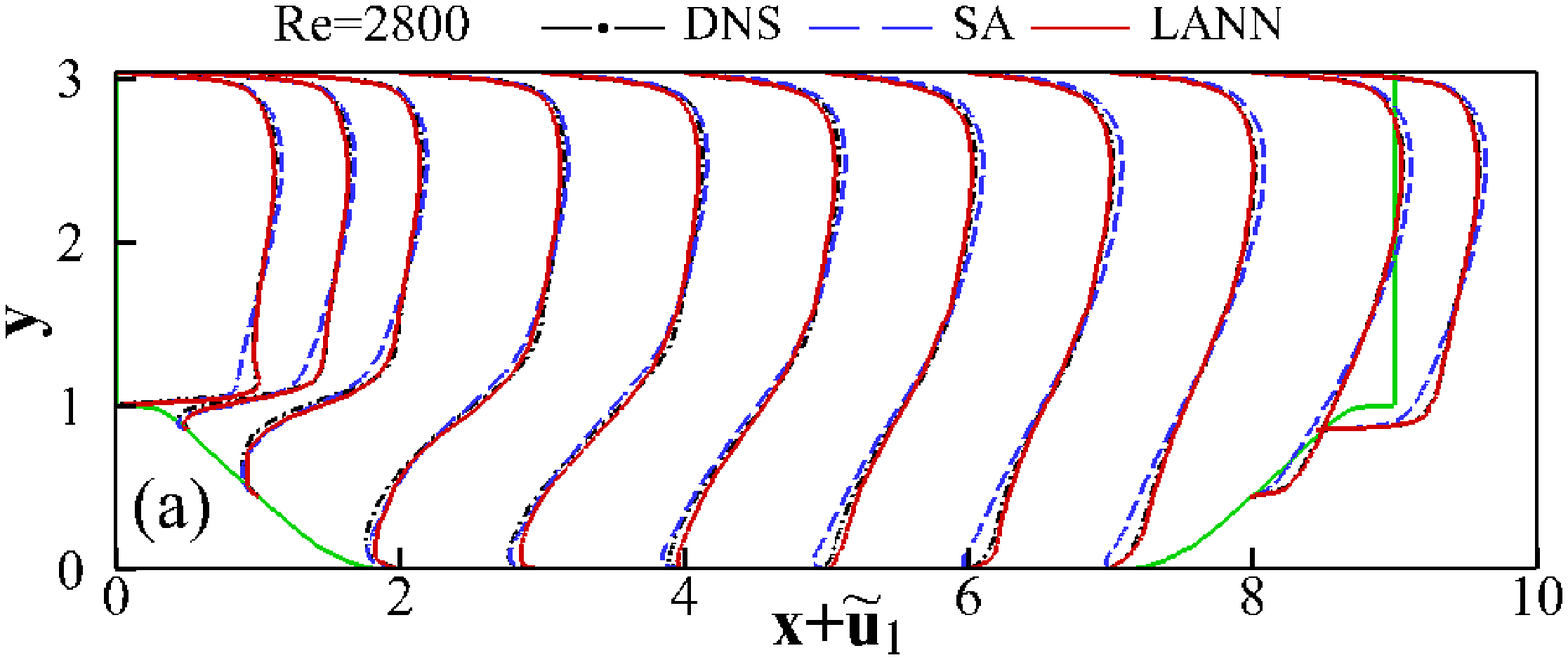}
\includegraphics[width=.8\textwidth]{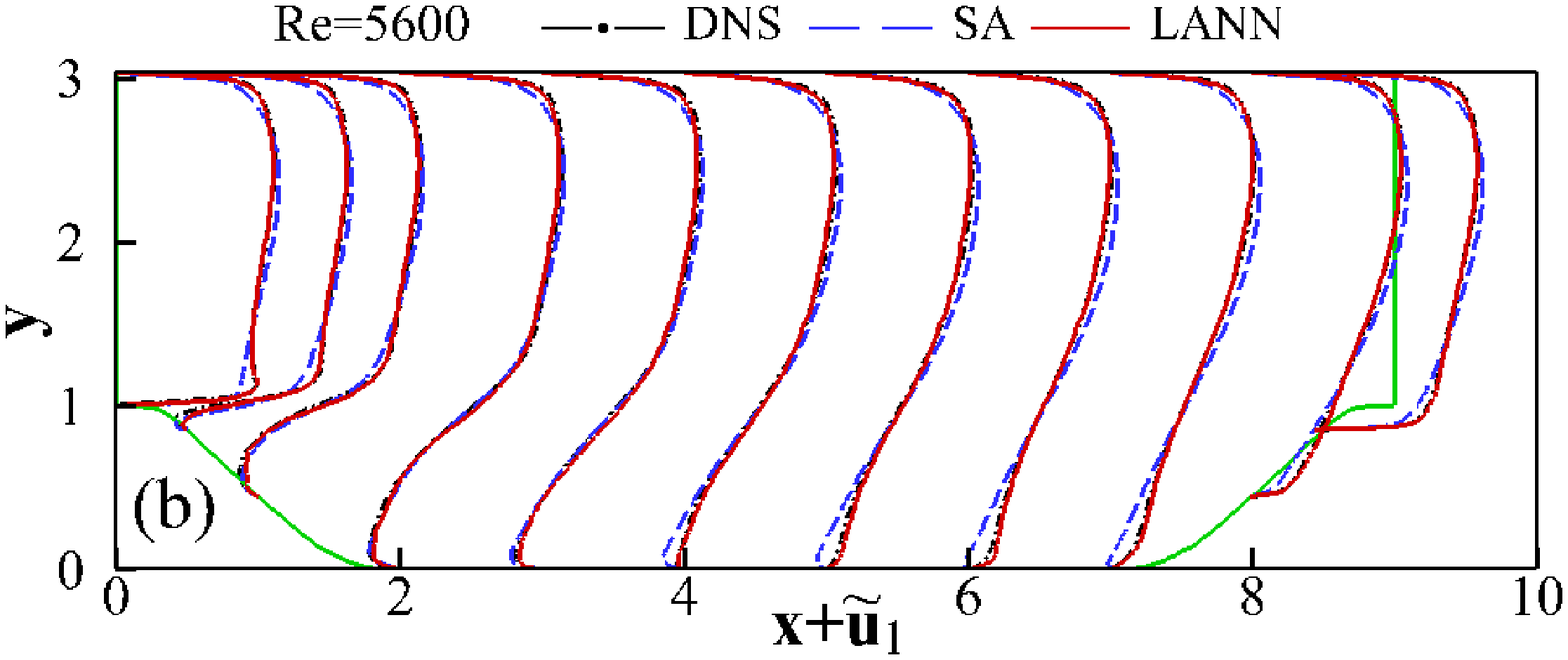}
\includegraphics[width=.8\textwidth]{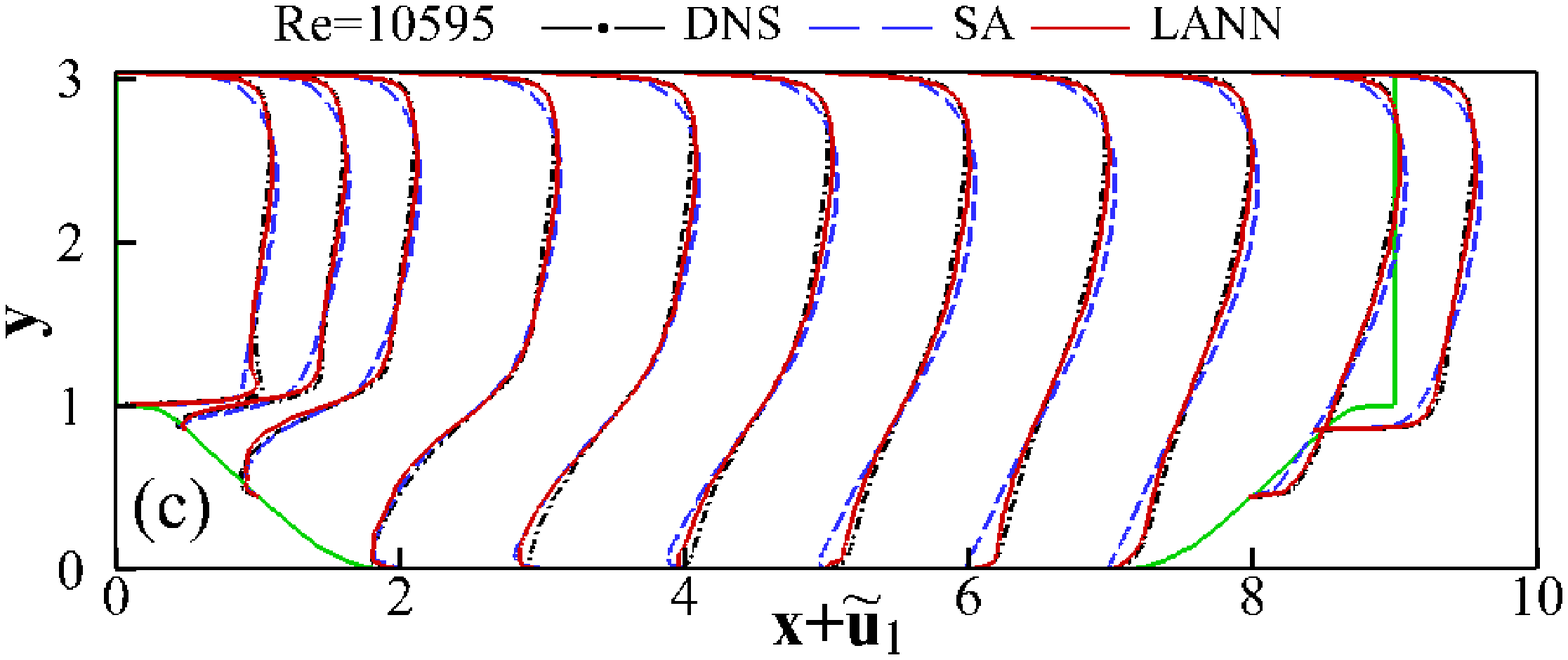}
 \caption{Mean streamwise velocity $\tilde{u}_{1}$ profiles with a grid resolution of $256\times129$ and $L_{x}=9.0$: (a) $Re=2800$, (b) $Re=5600$, (c) $Re=10595$.}\label{12}
\end{figure}

\begin{table}[tbp]
\centering
\begin{tabular}{lccccccccccc}
\hline\hline
$E_{r}\backslash x$ &0 &0.5 &1 &2 &3 &4 &5 &6 &7 &8 &8.5\\ \hline
SA   &0.035 &0.030 &0.026 &0.020 &0.026 &0.039  &0.054 &0.070 &0.061 &0.050 &0.044\\
LANN &0.012 &0.015 &0.018 &0.028 &0.029 &0.026  &0.021 &0.018 &0.015 &0.011 &0.010\\ \hline\hline
\end{tabular}
\caption{Relative error($E_{r}$) of $\tilde{u}_{1}$ for different models at $Re=2800$ and $y=2.5$ with the grid resolution of $256\times129$.}
\end{table}

Figure~\ref{11}(a) shows the distribution of the averaged wall shear stress $\tilde{\tau}_{w}$ along the lower wall at $Re=2800$ ($\tilde{\tau}_{w}=\frac{1}{Re}\frac{\partial \tilde{u}_{\xi}}{\partial \eta}$, where $\tilde{u}_{\xi}$ is the flow velocity parallel to the wall and $\eta$ is the distance to the wall). $\tilde{\tau}_{w}$ predicted by the LANN and SA models show similar behaviors as the DNS data. The peak of the profile of $\tilde{\tau}_{w}$ is recovered more accurately by the LANN model than the SA model. The averaged pressure along the lower wall at $Re=2800$ is shown in Fig.~\ref{11}(b). The mean pressure predicted by the LANN model is closer to the DNS data than that predicted by the SA model in the range $1\leq x\leq 8$. We compare the mean streamwise velocity $\tilde{u}_{1}$ at eleven locations in Fig.~\ref{12}. Both the SA and LANN models accurately predict $\tilde{u}_{1}$ near the upper wall. The RANS simulation with the SA model do a poor job near the lower wall, especially behind the separation. In contrast, $\tilde{u}_{1}$ predicted by the RANS simulation with the LANN model are in good agreement with the DNS data at all locations, suggesting that the LANN model can predict the mean streamwise velocity $\tilde{u}_{1}$ of flows over periodic hill accurately. Furthermore, the performance of the LANN model trained at $Re=2800$ are examined by predicting the mean streamwise velocity $\tilde{u}_{1}$ profile for flows over periodic hills with higher Reynolds numbers $Re=5600$ and $10595$. We display the mean streamwise velocities $\tilde{u}_{1}$ of DNS and RANS simulations with the SA and LANN models at $Re=5600$ and $10595$ in Fig.~\ref{12}(b-c). One can see that some errors occur near the lower and upper walls in the predicted $\tilde{u}_{1}$ by the SA model. In contrast, the results of the LANN model are very close to those of the DNS. Table~III shows relative errors of $\tilde{u}_{1}$ for different models at $Re=2800$ and $y=2.5$. We can see that the relative error of the LANN model is smaller than that of the SA model. Thus, the LANN model shows significant advantage over the SA model on relative error in the \emph{a posteriori} test.

\begin{figure}\centering
\includegraphics[width=.8\textwidth]{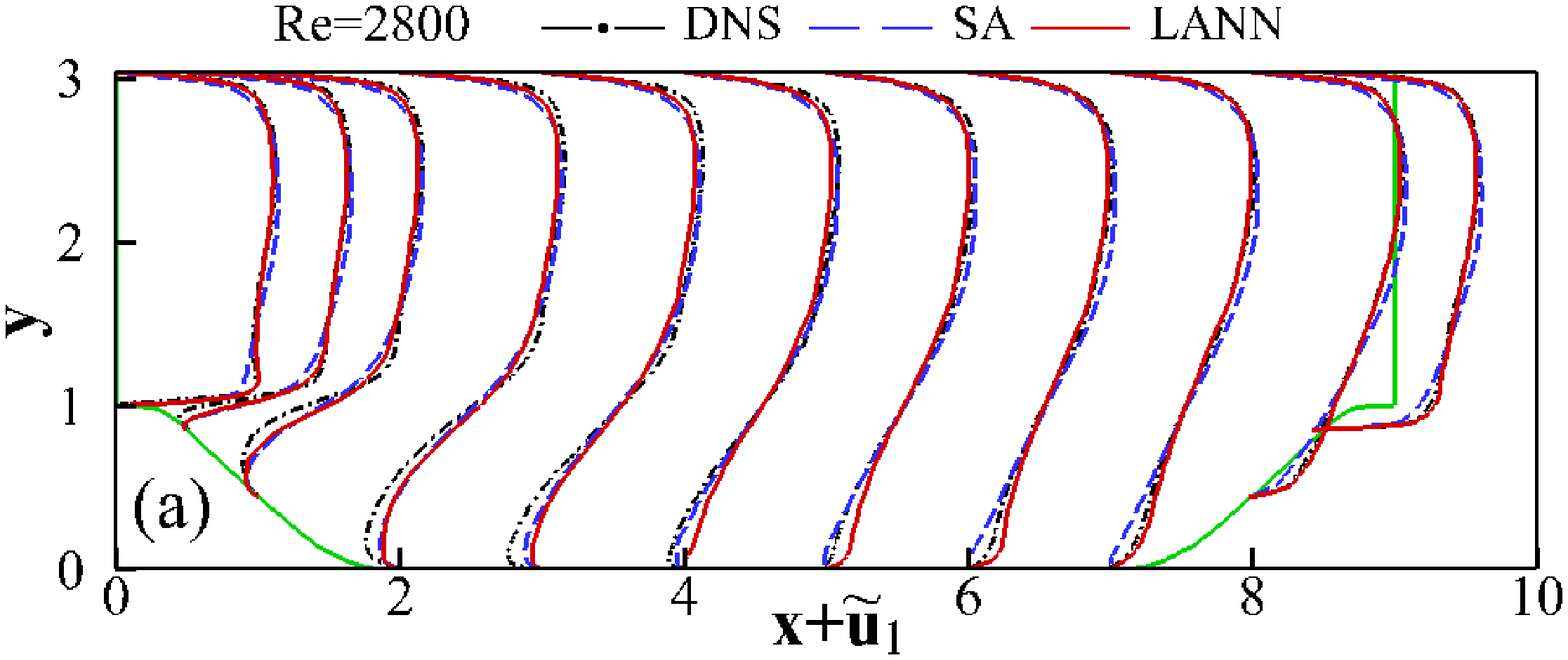}
\includegraphics[width=.8\textwidth]{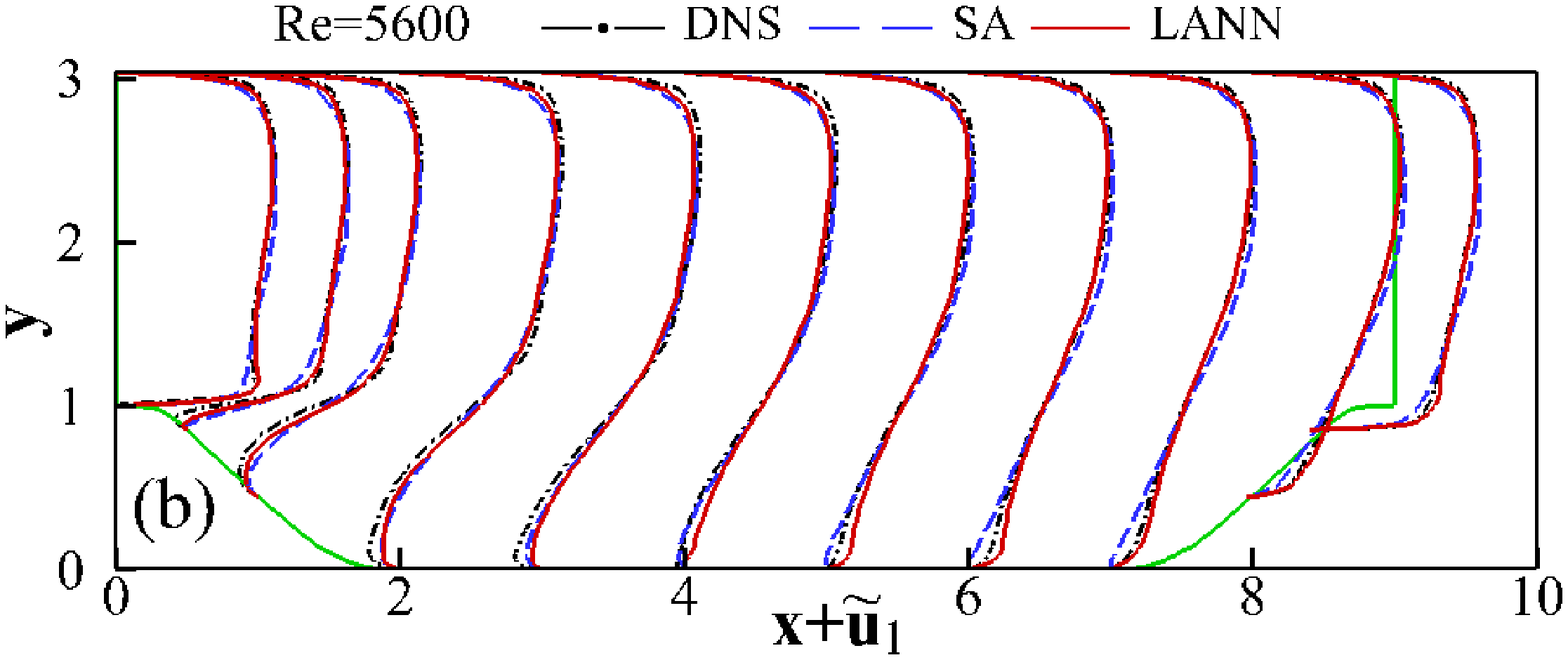}
\includegraphics[width=.8\textwidth]{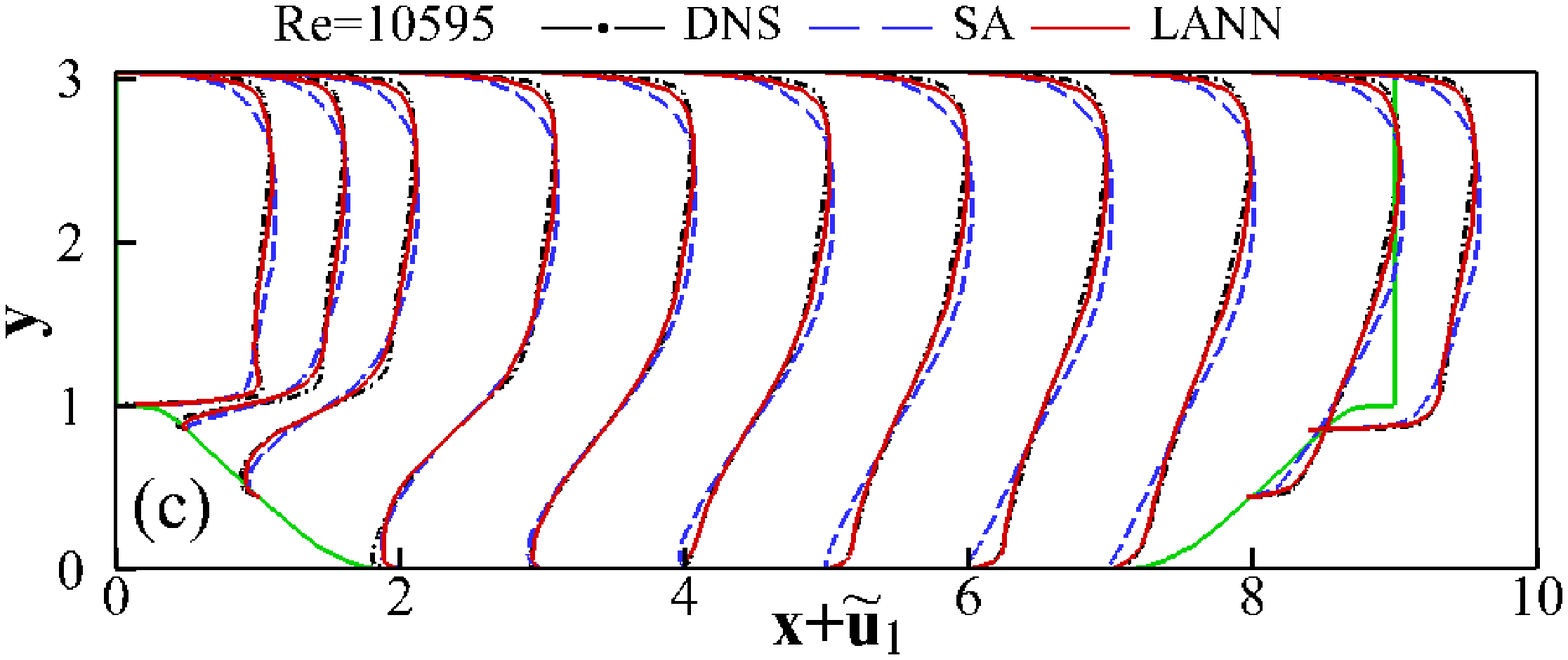}
 \caption{Mean streamwise velocity $\tilde{u}_{1}$ profiles with a grid resolution of $128\times65$ and $L_{x}=9.0$: (a) $Re=2800$, (b) $Re=5600$, (c) $Re=10595$.}\label{13}
\end{figure}

\begin{figure}\centering
\includegraphics[width=.8\textwidth]{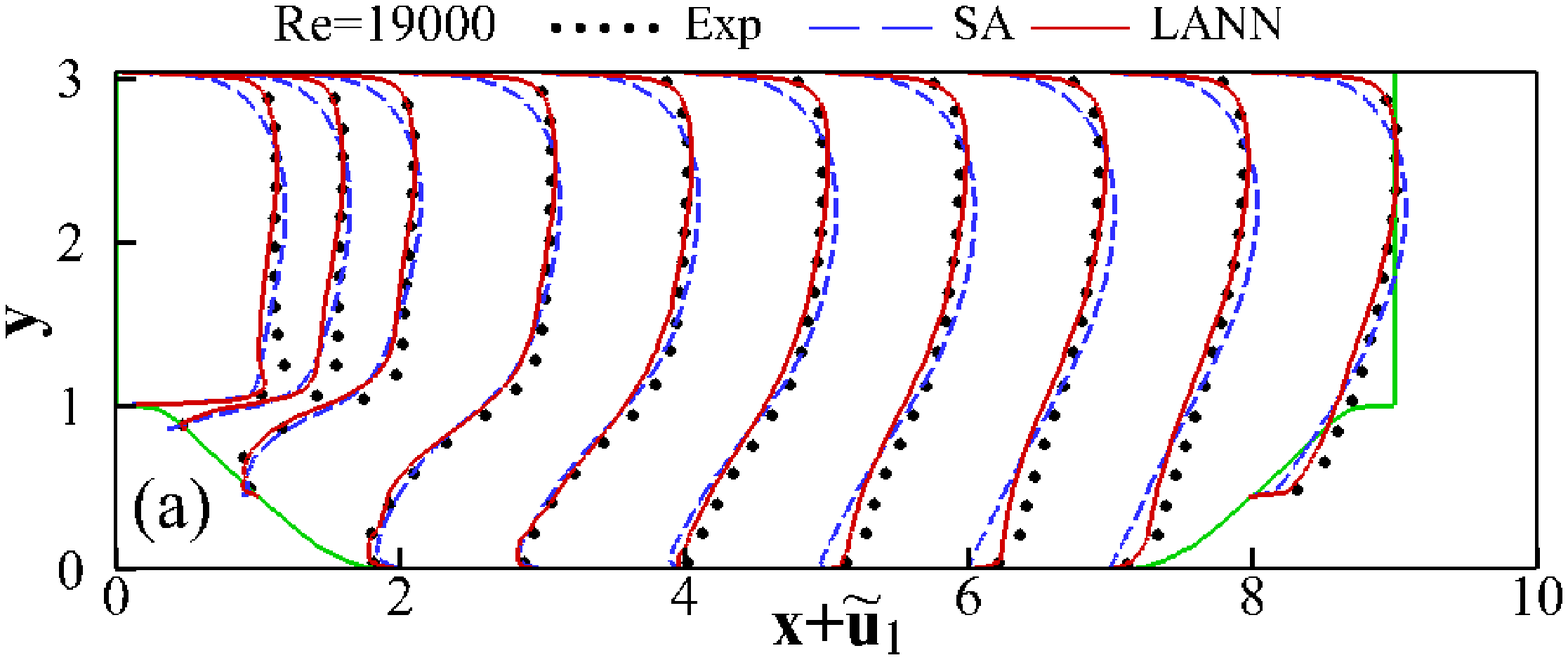}
\includegraphics[width=.8\textwidth]{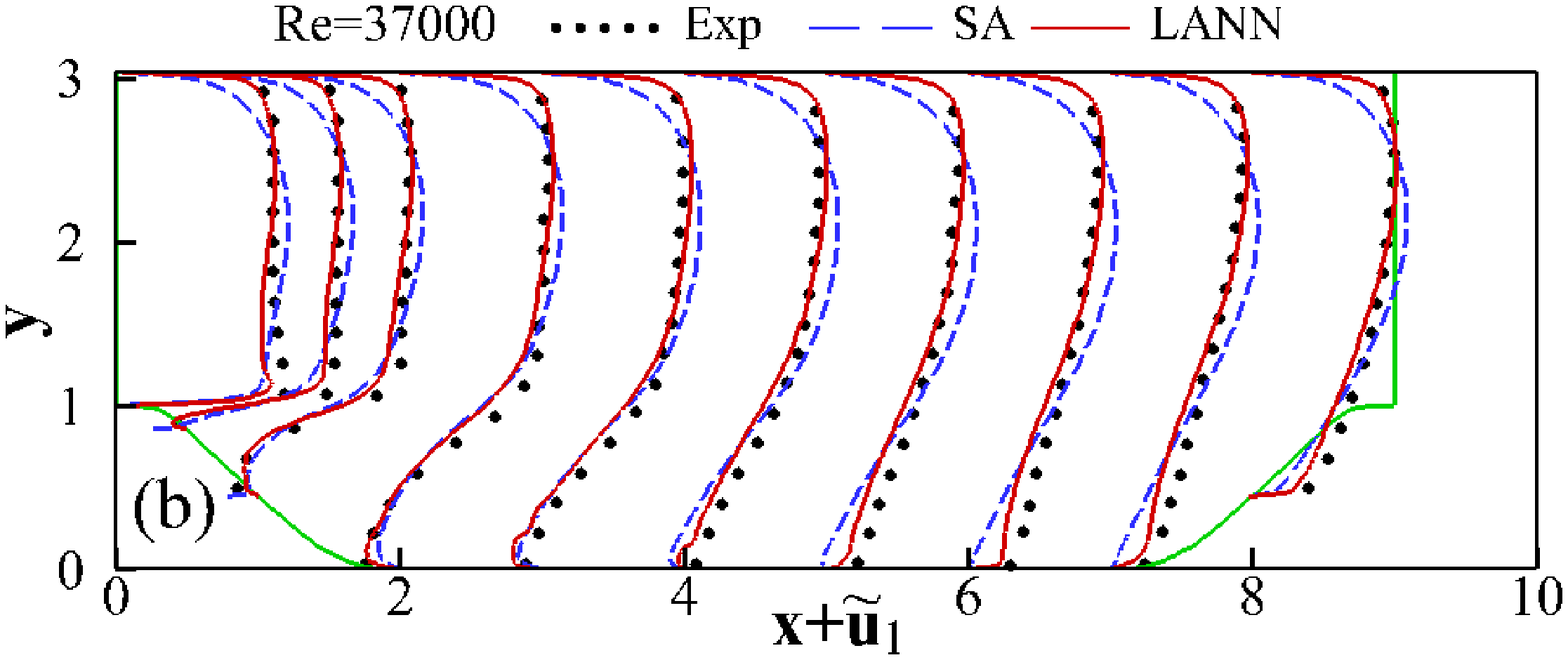}
 \caption{Mean streamwise velocity $\tilde{u}_{1}$ profiles with a grid resolution of $256\times129$ and $L_{x}=9.0$, ($\cdot\cdot\cdot\cdot\cdot$) experiments Rapp $\&$ Manhart\cite{Rapp2011}: a) $Re=19000$, (b) $Re=37000$.}\label{14}
\end{figure}

\begin{figure}\centering
\includegraphics[width=.8\textwidth]{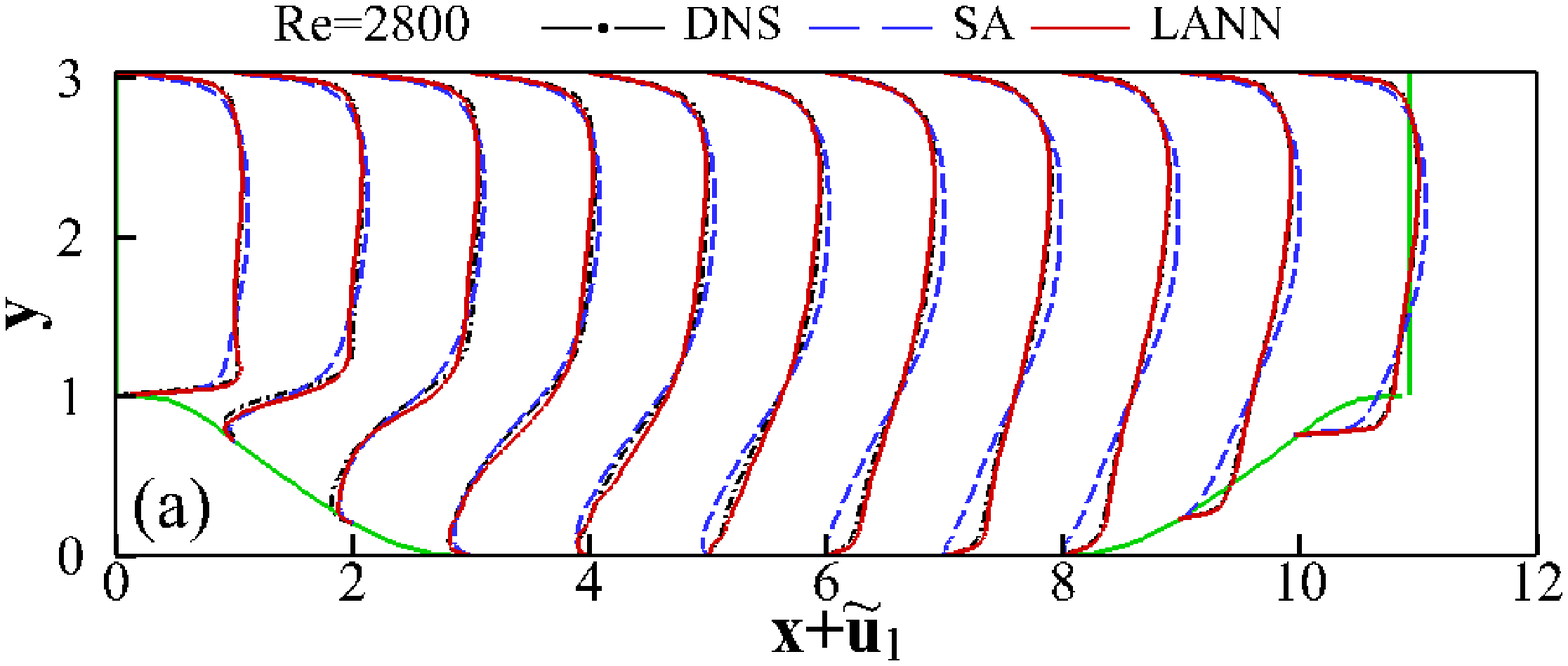}
\includegraphics[width=.8\textwidth]{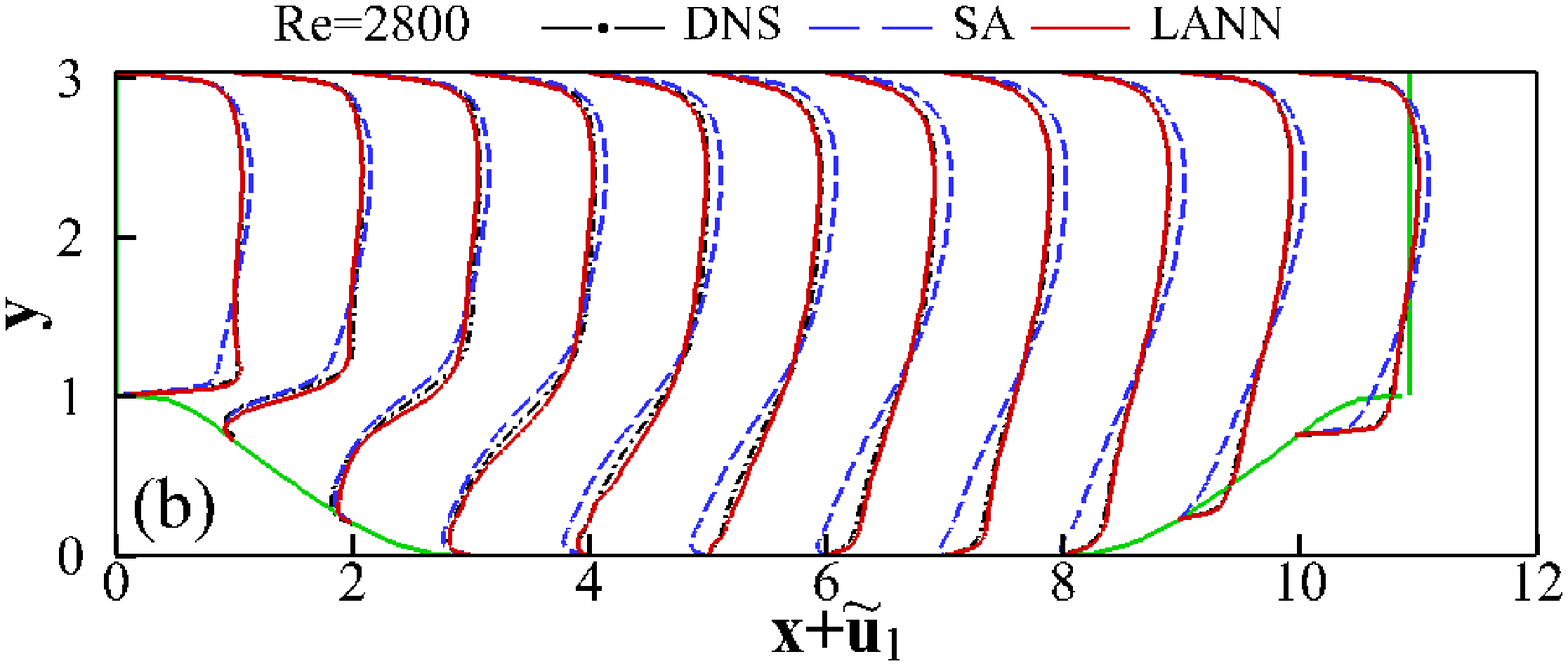}
 \caption{Mean streamwise velocity $\tilde{u}_{1}$ profiles at $Re=2800$ with $L_{x}=10.929$ ($\alpha=1.5$): (a) grid resolution of $128\times65$, (b) grid resolution of $256\times129$.}\label{15}
\end{figure}

The generality of the LANN model are examined by plotting the mean streamwise velocity $\tilde{u}_{1}$ profiles for flows over periodic hills on coarser grids and in a different computational domain (the detailed information about the new $L_{x}$ can be given with $L_{x}=3.858\alpha+5.142$, where $\alpha=1.5$)\cite{Xiao2020}. The mean streamwise velocity $\tilde{u}_{1}$ profiles for $Re=2800$, 5600, 10595 with a grid resolution of $128\times65$ are shown in Fig.~\ref{13}. $\tilde{u}_{1}$ predicted by the LANN model is much closer to those of DNS, compared to those predicted by the SA model. Furthermore, we evaluated the performance of the LANN model against the experimental results by Rapp \emph{et al.} at $Re=19000$ and $37000$\cite{Rapp2011}. As shown in Fig.~\ref{14}, compared with the results from the experiments\cite{Rapp2011}, the mean streamwise velocity $\tilde{u}_{1}$ from the LANN model match better than those from the SA model. Finally, figure~\ref{15} shows the mean streamwise velocity $\tilde{u}_{1}$ for flow over periodic hill with a total horizontal length of $L_{x}=10.929$ ($\alpha=1.5$). The results from the LANN model show good agreement with the DNS results, which are better than those obtained with the SA model under the same grid resolutions.

\section{Discussion}
The flow over periodic hill is one of the standard examples for developing new turbulence models of RANS\cite{Duraisamy2019}, which includes the separation, recirculation, and reattachment. One important characteristic of RANS is that the RANS unclosed terms are very complex in turbulence near the boundary. It's hard to reconstruct the RANS unclosed terms accurately and stably near a wall, which depend stronly on the distance to the walls\cite{Spalart1994,Duraisamy2019}. Due to the irregular and diverse nature of turbulence, it is difficult to explicitly derive the dependence of RANS unclosed terms on the mean flow properties with analytical methods. The advantage of the LANN model in the local reference frame is that the $\eta$ axis of the local coordinate system is orthogonal to the nearer wall and the nearest distance from the walls $d$ can be measured along the coordinate axis $\eta$, which is general and versatile for complex wall conditions. In this research, it has been demonstrated that the LANN method is a powerful tool which can efficiently learn the high-dimensional and nonlinear relations between the RANS unclosed terms and the mean flow fields for flows over periodic hills with varying slopes. The effects of more complex boundary conditions on the RANS simulations of wall-bounded turbulent flows will be modeled with the LANN framework in a follow-up study.

\section{Conclusions}

In this work, we proposed a framework of LANN for the RANS unclosed terms in RANS simulations of compressible turbulence. The proposed LANN model depends on the local coordinate system orthogonal to the wall for flows over periodic hills. In the \emph{a priori} test, the correlation coefficients are larger than 0.96 and the relative errors are smaller than 18 $\%$ for the LANN model. In an \emph{a posteriori} analysis, we compare the performances of the LANN model with those of the SA model in the predictions of the average velocity, wall-shear stress and average pressure in flows over periodic hills with varying slopes $\alpha=1$ and $1.5$. There are non-negligible errors between the mean velocities predicted by the SA model and the results of DNS near the walls, especially in the region right behind the separation. In contrast, the LANN model predicts the mean velocity accurately, and it also reconstruct the mean pressure closer to those of the DNS than those with the SA model at Reynolds numbers $Re=2800$, 5600, 10595, 19000, 37000. In addition, the mean velocity in flows over periodic hills with longer horizontal width $L_{x}=10.929$ ($\alpha=1.5$) predicted by the LANN model is in well agreement with that of the DNS. The above comparison showed that the LANN model outperformed the SA model in the flows over periodic hills.

The LANN model should also be very useful to wall-bounded turbulent flows with curved walls. There are several issues that need further exploration: the physical relationship between averaged flow fields and the RANS unclosed terms, the hyperparameter space, the symmetry and interpretation of the neural network models, the non-locality characteristics of the RANS dynamics, and applications in more complex flows.

\begin{acknowledgments}
We thank Weinan E and Chao Ma for helpful discussions. The authors are grateful to Xinliang Li for providing the CFD codes OpenCFD-SC and OpenCFD-EC. The work of Chenyue Xie is supported in part by a gift to Princeton University from iFlytek. The work of Jianchun Wang is supported by the National Natural Science Foundation of China (NSFC Grants No. 91952104).

\end{acknowledgments}

\begin{thebibliography}{100}
\expandafter\ifx\csname
natexlab\endcsname\relax\def\natexlab#1{#1}\fi
\expandafter\ifx\csname bibnamefont\endcsname\relax
  \def\bibnamefont#1{#1}\fi
\expandafter\ifx\csname bibfnamefont\endcsname\relax
  \def\bibfnamefont#1{#1}\fi
\expandafter\ifx\csname citenamefont\endcsname\relax
  \def\citenamefont#1{#1}\fi
\expandafter\ifx\csname url\endcsname\relax
  \def\url#1{\texttt{#1}}\fi
\expandafter\ifx\csname
urlprefix\endcsname\relax\def\urlprefix{URL }\fi
\providecommand{\bibinfo}[2]{#2}
\providecommand{\eprint}[2][]{\url{#2}}


\bibitem[{\citenamefont{Speziale et~al.}(1991)\citenamefont{Speziale}}]{Speziale1991}
\bibinfo{author}{\bibfnamefont{C. G.} \bibnamefont{Speziale}},
\bibinfo{title}{Analytical methods for the development of Reynodls-stress closures in turbulence},
\bibinfo{journal}{Annu. Rev. Fluid Mech.}
\textbf{\bibinfo{volume}{23}}, \bibinfo{pages}{107-157}
(\bibinfo{year}{1991}).

\bibitem[{\citenamefont{Pope et~al.}(2000)\citenamefont{Pope}}]{Pope2000}
\bibinfo{author}{\bibfnamefont{S. B.}~\bibnamefont{Pope}},
\bibinfo{book}{{\it Turbulent Flows}}
(\bibinfo{year}{Cambridge University Press, 2000}).

\bibitem[{\citenamefont{Durbin et~al.}(2018)\citenamefont{Durbin}}]{Durbin2018}
\bibinfo{author}{\bibfnamefont{P. A.} \bibnamefont{Durbin}},
\bibinfo{title}{Some recent developments in turbulence closure modeling},
\bibinfo{journal}{Annu. Rev. Fluid Mech.}
\textbf{\bibinfo{volume}{50}}, \bibinfo{pages}{77-103}
(\bibinfo{year}{2018}).

\bibitem[{\citenamefont{Reynolds et~al.}(1895)\citenamefont{Reynolds}}]{Reynolds1895}
\bibinfo{author}{\bibfnamefont{O.} \bibnamefont{Reynolds}},
\bibinfo{title}{On the dynamical theory of incompressible viscous fluids and the determination of the criterion},
\bibinfo{journal}{Philos. Trans. R. Soc. Lond. A}
\textbf{\bibinfo{volume}{186}}, \bibinfo{pages}{123-164}
(\bibinfo{year}{1895}).

\bibitem[{\citenamefont{Boussinesq et~al.}(1877)\citenamefont{Boussinesq}}]{Boussinesq1877}
\bibinfo{author}{\bibfnamefont{J.} \bibnamefont{Boussinesq}},
\bibinfo{title}{Th$\acute{e}$orie de l'$\acute{e}$coulement tourbillant},
\bibinfo{journal}{Pr$\acute{e}$sent$\acute{e}$s par divers Savants Acad. Sci. Inst. Fr.}
\textbf{\bibinfo{volume}{23}}, \bibinfo{pages}{46-50}
(\bibinfo{year}{1877}).

\bibitem[{\citenamefont{Prandtl et~al.}(1925)\citenamefont{Prandtl}}]{Prandtl1925}
\bibinfo{author}{\bibfnamefont{L.} \bibnamefont{Prandtl}},
\bibinfo{title}{\"{U}ber die ausgebildete Turbulenz},
\bibinfo{journal}{ZAMM}
\textbf{\bibinfo{volume}{5}}, \bibinfo{pages}{136-139}
(\bibinfo{year}{1925}).

\bibitem[{\citenamefont{Spalart et~al.}(1994)\citenamefont{Spalart}}]{Spalart1994}
\bibinfo{author}{\bibfnamefont{P. R.} \bibnamefont{Spalart}},
\bibnamefont{and} \bibinfo{author}{\bibfnamefont{S. R.}~\bibnamefont{Allmaras}},
\bibinfo{title}{A One-Equation Turbulence Model for Aerodynamic Flows},
\bibinfo{journal}{Rech. Aerosp.}
\textbf{\bibinfo{volume}{1}}, \bibinfo{pages}{5-21}
(\bibinfo{year}{1994}).

\bibitem[{\citenamefont{Jones et~al.}(1972)\citenamefont{Jones}}]{Jones1972}
\bibinfo{author}{\bibfnamefont{W. P.} \bibnamefont{Jones}},
\bibnamefont{and} \bibinfo{author}{\bibfnamefont{B. E.}~\bibnamefont{Launder}},
\bibinfo{title}{The Prediction of Laminarization with a Two-Equation Model of Turbulence},
\bibinfo{journal}{Int. J. Heat Mass Transf.}
\textbf{\bibinfo{volume}{15}}, \bibinfo{pages}{301-314}
(\bibinfo{year}{1972}).

\bibitem[{\citenamefont{Launder et~al.}(1974)\citenamefont{Launder}}]{Launder1974}
\bibinfo{author}{\bibfnamefont{B. E.} \bibnamefont{Launder}},
\bibnamefont{and} \bibinfo{author}{\bibfnamefont{B. I.}~\bibnamefont{Sharma}},
\bibinfo{title}{Application of the Energy Dissipation Model of Turbulence to the Calculation of Flow Near a Spinning Disc},
\bibinfo{journal}{Lett. Heat Mass Transf.}
\textbf{\bibinfo{volume}{1}}, \bibinfo{pages}{131-138}
(\bibinfo{year}{1974}).

\bibitem[{\citenamefont{Yakhot et~al.}(1986)\citenamefont{Yakhot}}]{Yakhot1986}
\bibinfo{author}{\bibfnamefont{V.} \bibnamefont{Yakhot}},
\bibnamefont{and} \bibinfo{author}{\bibfnamefont{S. A.}~\bibnamefont{Orszag}},
\bibinfo{title}{Renormalization group analysis of turbulence. I. Basic theory},
\bibinfo{journal}{J. Sci. Comput.}
\textbf{\bibinfo{volume}{1}}, \bibinfo{pages}{3-51}
(\bibinfo{year}{1986}).

\bibitem[{\citenamefont{Yakhot et~al.}(1992)\citenamefont{Yakhot}}]{Yakhot1992}
\bibinfo{author}{\bibfnamefont{V.} \bibnamefont{Yakhot}},
\bibinfo{author}{\bibfnamefont{S. A.} \bibnamefont{Orszag}},
\bibinfo{author}{\bibfnamefont{S.} \bibnamefont{Thangam}},
\bibinfo{author}{\bibfnamefont{T. B.} \bibnamefont{Gatski}},
\bibnamefont{and} \bibinfo{author}{\bibfnamefont{C. G.}~\bibnamefont{Speziale}},
\bibinfo{title}{Development of turbulence models for shear flows by a double expansion technique},
\bibinfo{journal}{Phys. Fluids A Fluid Dyn.}
\textbf{\bibinfo{volume}{4}}, \bibinfo{pages}{1510-1520}
(\bibinfo{year}{1992}).

\bibitem[{\citenamefont{Menter et~al.}(1994)\citenamefont{Menter}}]{Menter1994}
\bibinfo{author}{\bibfnamefont{F. R.} \bibnamefont{Menter}},
\bibinfo{title}{Two-equation eddy-viscosity turbulence models for engineering applications},
\bibinfo{journal}{AIAA J.}
\textbf{\bibinfo{volume}{32}}, \bibinfo{pages}{1598-1605}
(\bibinfo{year}{1994}).

\bibitem[{\citenamefont{Spalart et~al.}(2000)\citenamefont{Spalart}}]{Spalart2000}
\bibinfo{author}{\bibfnamefont{P. R.} \bibnamefont{Spalart}},
\bibinfo{title}{Strategies for turbulence modeling and simulations},
\bibinfo{journal}{Int. J. Heat Fluid Flow}
\textbf{\bibinfo{volume}{21}}, \bibinfo{pages}{252-263}
(\bibinfo{year}{2000}).

\bibitem[{\citenamefont{Wilcox et~al.}(2008)\citenamefont{Wilcox}}]{Wilcox2008}
\bibinfo{author}{\bibfnamefont{D. C.} \bibnamefont{Wilcox}},
\bibinfo{title}{Formulation of the $k-\omega$ turbulence model revisited},
\bibinfo{journal}{AIAA J.}
\textbf{\bibinfo{volume}{46}}, \bibinfo{pages}{2823-2838}
(\bibinfo{year}{2008}).

\bibitem[{\citenamefont{Launder et~al.}(1975)\citenamefont{Launder}}]{Launder1975}
\bibinfo{author}{\bibfnamefont{B. E.} \bibnamefont{Launder}},
\bibinfo{author}{\bibfnamefont{G. J.} \bibnamefont{Reece}},
\bibnamefont{and} \bibinfo{author}{\bibfnamefont{W.}~\bibnamefont{Rodi}},
\bibinfo{title}{Progress in the Development of a Reynolds-Stress Turbulent Closure},
\bibinfo{journal}{J. Fluid Mech.}
\textbf{\bibinfo{volume}{68(3)}}, \bibinfo{pages}{537-566}
(\bibinfo{year}{1975}).

\bibitem[{\citenamefont{Tracey et~al.}(2013)\citenamefont{Tracey}}]{Tracey2013}
\bibinfo{author}{\bibfnamefont{B. D.} \bibnamefont{Tracey}},
\bibinfo{author}{\bibfnamefont{K.} \bibnamefont{Duraisamy}},
\bibnamefont{and} \bibinfo{author}{\bibfnamefont{J. J.}~\bibnamefont{Alonso}},
\bibinfo{title}{Application of supervised learning to quantify uncertianties in turbulence and combustion modeling},
\bibinfo{journal}{51rd AIAA Aerospace Sciences Meeting, Dallas, TX}
\textbf{\bibinfo{volume}{}}, \bibinfo{pages}{2013-0259}
(\bibinfo{year}{2013}).

\bibitem[{\citenamefont{Duraisamy et~al.}(2015)\citenamefont{Duraisamy}}]{Duraisamy2015}
\bibinfo{author}{\bibfnamefont{K.} \bibnamefont{Duraisamy}},
\bibinfo{author}{\bibfnamefont{Z. J.} \bibnamefont{Zhang}},
\bibnamefont{and} \bibinfo{author}{\bibfnamefont{A. P.}~\bibnamefont{Singh}},
\bibinfo{title}{New approaches in turbulence and transition modeling using data-driven techniques},
\bibinfo{journal}{Proc. 53rd AIAA Aerospace Sciencs Meeting}
\textbf{\bibinfo{volume}{}}, \bibinfo{pages}{2015-1284}
(\bibinfo{year}{2015}).

\bibitem[{\citenamefont{Parish et~al.}(2016)\citenamefont{Parish}}]{Parish2016}
\bibinfo{author}{\bibfnamefont{E. J.} \bibnamefont{Parish}},
\bibnamefont{and} \bibinfo{author}{\bibfnamefont{K.}~\bibnamefont{Duraisamy}},
\bibinfo{title}{A paradigm for data-driven predictive modeling using field inversion and machine learning},
\bibinfo{journal}{J. Comput. Phys.}
\textbf{\bibinfo{volume}{305}}, \bibinfo{pages}{758-774}
(\bibinfo{year}{2016}).

\bibitem[{\citenamefont{Weatheritt et~al.}(2016)\citenamefont{Weatheritt}}]{Weatheritt2016}
\bibinfo{author}{\bibfnamefont{J.} \bibnamefont{Weatheritt}},
\bibnamefont{and} \bibinfo{author}{\bibfnamefont{R.}~\bibnamefont{Sandberg}},
\bibinfo{title}{A novel evolutionary algorithm applied to algebraic modifications of the RANS stresscstrain relationship},
\bibinfo{journal}{J. Comput. Phys.}
\textbf{\bibinfo{volume}{325}}, \bibinfo{pages}{22-37}
(\bibinfo{year}{2016}).


\bibitem[{\citenamefont{Ling et~al.}(2015)\citenamefont{Ling}}]{Ling2015}
\bibinfo{author}{\bibfnamefont{J.} \bibnamefont{Ling}},
\bibnamefont{and} \bibinfo{author}{\bibfnamefont{J.}~\bibnamefont{Templeton}},
\bibinfo{title}{Evaluation of machine learning algorithms for prediction of regions of high Reynolds averaged Navier Stokes uncertainty},
\bibinfo{journal}{Phys. Fluids}
\textbf{\bibinfo{volume}{27}}, \bibinfo{pages}{085103}
(\bibinfo{year}{2015}).

\bibitem[{\citenamefont{Zhang et~al.}(2015)\citenamefont{Zhang}}]{Zhang2015}
\bibinfo{author}{\bibfnamefont{Z. J.} \bibnamefont{Zhang}},
\bibnamefont{and} \bibinfo{author}{\bibfnamefont{K.}~\bibnamefont{Duraisamy}},
\bibinfo{title}{Machine learning methods for data-driven turbulence modeling},
\bibinfo{journal}{Proceedings of the 22nd AIAA Computational Fluid Dynamics Conference (AIAA, Reston)}
\textbf{\bibinfo{volume}{}}, \bibinfo{pages}{2015-2460}
(\bibinfo{year}{2015}).

\bibitem[{\citenamefont{Ling et~al.}(2016)\citenamefont{Ling}}]{Ling2016}
\bibinfo{author}{\bibfnamefont{J.} \bibnamefont{Ling}},
\bibinfo{author}{\bibfnamefont{A.}~\bibnamefont{Kurzawski}},
\bibnamefont{and} \bibinfo{author}{\bibfnamefont{J.}~\bibnamefont{Templeton}},
\bibinfo{title}{Reynolds averaged turbulence modelling using deep neural networks with embedded invariance},
\bibinfo{journal}{J. Fluid Mech.}
\textbf{\bibinfo{volume}{807}}, \bibinfo{pages}{155}
(\bibinfo{year}{2016}).

\bibitem[{\citenamefont{Ling et~al.}(2016)\citenamefont{Ling}}]{Ling2016b}
\bibinfo{author}{\bibfnamefont{J.} \bibnamefont{Ling}},
\bibinfo{author}{\bibfnamefont{R.} \bibnamefont{Jones}},
\bibnamefont{and} \bibinfo{author}{\bibfnamefont{J.}~\bibnamefont{Templeton}},
\bibinfo{title}{Machine learning strategies for systems with invariance properties},
\bibinfo{journal}{J. Compt. Phys.}
\textbf{\bibinfo{volume}{318}}, \bibinfo{pages}{22}
(\bibinfo{year}{2016}).

\bibitem[{\citenamefont{Xiao et~al.}(2016)\citenamefont{Xiao}}]{Xiao2016}
\bibinfo{author}{\bibfnamefont{H.} \bibnamefont{Xiao}},
\bibinfo{author}{\bibfnamefont{J. L.}~\bibnamefont{Wu}},
\bibinfo{author}{\bibfnamefont{J. X.} \bibnamefont{Wang}},
\bibinfo{author}{\bibfnamefont{J. X.}~\bibnamefont{Sun}},
\bibnamefont{and} \bibinfo{author}{\bibfnamefont{C. J. R.}~\bibnamefont{Roy}},
\bibinfo{title}{Quantifying and reducing model-form uncertainties in Reynolds-averaged Navier-Stokes simulations: a data-driven, physics-informed Bayesian approach},
\bibinfo{journal}{J. Comput. Phys.}
\textbf{\bibinfo{volume}{324}}, \bibinfo{pages}{115}
(\bibinfo{year}{2016}).

\bibitem[{\citenamefont{Tracey et~al.}(2015)\citenamefont{Tracey}}]{Tracey2015}
\bibinfo{author}{\bibfnamefont{B. D.} \bibnamefont{Tracey}},
\bibinfo{author}{\bibfnamefont{K.} \bibnamefont{Duraisamy}},
\bibnamefont{and} \bibinfo{author}{\bibfnamefont{J. J.}~\bibnamefont{Alonso}},
\bibinfo{title}{A machine learning strategy to assist turbulence model development},
\bibinfo{journal}{53rd AIAA Aerospace Sciences Meeting, Kissimmee, FL}
\textbf{\bibinfo{volume}{}}, \bibinfo{pages}{2015-1287}
(\bibinfo{year}{2015}).

\bibitem[{\citenamefont{Brunton et~al.}(2016)\citenamefont{Brunton}}]{Brunton2016}
\bibinfo{author}{\bibfnamefont{S. L.} \bibnamefont{Brunton}},
\bibinfo{author}{\bibfnamefont{J. L.} \bibnamefont{Proctor}},
\bibnamefont{and} \bibinfo{author}{\bibfnamefont{J. N.}~\bibnamefont{Kutz}},
\bibinfo{title}{Discovering governing equations from data by sparse identification of nonlinear dynamical systems},
\bibinfo{journal}{Proc. Natl Acad. Sci. USA}
\textbf{\bibinfo{volume}{113}}, \bibinfo{pages}{3932}
(\bibinfo{year}{2016}).

\bibitem[{\citenamefont{Wang et~al.}(2017)\citenamefont{Wang}}]{Wangj2017}
\bibinfo{author}{\bibfnamefont{J. X.} \bibnamefont{Wang}},
\bibinfo{author}{\bibfnamefont{J. L.} \bibnamefont{Wu}},
\bibnamefont{and} \bibinfo{author}{\bibfnamefont{H.}~\bibnamefont{Xiao}},
\bibinfo{title}{Physics-informed machine learning approach for reconstructing reynolds stress modeling discrepancies based on dns data},
\bibinfo{journal}{Phys. Rev. Fluids}
\textbf{\bibinfo{volume}{2}}, \bibinfo{pages}{034603}
(\bibinfo{year}{2017}).

\bibitem[{\citenamefont{Duraisamy et~al.}(2017)\citenamefont{Duraisamy}}]{Duraisamy2017}
\bibinfo{author}{\bibfnamefont{K.} \bibnamefont{Duraisamy}},
\bibinfo{author}{\bibfnamefont{A. P.} \bibnamefont{Singh}},
\bibnamefont{and} \bibinfo{author}{\bibfnamefont{S.}~\bibnamefont{Pan}},
\bibinfo{title}{Augmentation of turbulence models using field inversion and machine learning},
\bibinfo{journal}{55th AIAA Aerospace Sciences Meeting, (Grapevine, Texas)}
\textbf{\bibinfo{volume}{}}, \bibinfo{pages}{2017-0993}
(\bibinfo{year}{2017}).


\bibitem[{\citenamefont{Kutz et~al.}(2017)\citenamefont{Kutz}}]{Kutz2017}
\bibinfo{author}{\bibfnamefont{J. N.} \bibnamefont{Kutz}},
\bibinfo{title}{Deep learning in fluid dynamics},
\bibinfo{journal}{J. Fluid Mech.}
\textbf{\bibinfo{volume}{814}}, \bibinfo{pages}{1}
(\bibinfo{year}{2017}).

\bibitem[{\citenamefont{Vollant et~al.}(2017)\citenamefont{Vollant}}]{Vollant2017}
\bibinfo{author}{\bibfnamefont{A.} \bibnamefont{Vollant}},
\bibinfo{author}{\bibfnamefont{G.} \bibnamefont{Balarac}},
\bibnamefont{and} \bibinfo{author}{\bibfnamefont{C.}~\bibnamefont{Corre}},
\bibinfo{title}{Subgrid-scale scalar flux modelling based on optimal estimation theory and machine-learning procedures},
\bibinfo{journal}{J. Turbulence}
\textbf{\bibinfo{volume}{18(9)}}, \bibinfo{pages}{854}
(\bibinfo{year}{2017}).

\bibitem[{\citenamefont{Mohan et~al.}(2018)\citenamefont{Mohan}}]{Mohan2018}
\bibinfo{author}{\bibfnamefont{A. T.} \bibnamefont{Mohan}},
\bibnamefont{and} \bibinfo{author}{\bibfnamefont{D. V.}~\bibnamefont{Gaitonde}},
\bibinfo{title}{A deep learning based approach to reduced order modeling for turbulent flow control using LSTM neural networks},
\bibinfo{journal}{arXiv:1804.09269}
(\bibinfo{year}{2018}).

\bibitem[{\citenamefont{Raissi et~al.}(2018)\citenamefont{Raissi}}]{Raissi2018}
\bibinfo{author}{\bibfnamefont{M.} \bibnamefont{Raissi}},
\bibnamefont{and} \bibinfo{author}{\bibfnamefont{G. E.}~\bibnamefont{Karniadakis}},
\bibinfo{title}{Hidden physics models: machine learning of nonlinear partial differential equations},
\bibinfo{journal}{J. Comput. Phys.}
\textbf{\bibinfo{volume}{357}}, \bibinfo{pages}{125}
(\bibinfo{year}{2018}).

\bibitem[{\citenamefont{Wu et~al.}(2018)\citenamefont{Wu}}]{Wu2018}
\bibinfo{author}{\bibfnamefont{J. L.} \bibnamefont{Wu}},
\bibinfo{author}{\bibfnamefont{H.} \bibnamefont{Xiao}},
\bibnamefont{and} \bibinfo{author}{\bibfnamefont{E.}~\bibnamefont{Paterson}},
\bibinfo{title}{Physics-informed machine learning approach for augmenting turbulence models: A comprehensive framework},
\bibinfo{journal}{Phys. Rev. Fluids}
\textbf{\bibinfo{volume}{3}}, \bibinfo{pages}{074602}
(\bibinfo{year}{2018}).

\bibitem[{\citenamefont{Jofre et~al.}(2017)\citenamefont{Jofre}}]{Jofre2018}
\bibinfo{author}{\bibfnamefont{L.} \bibnamefont{Jofre}},
\bibinfo{author}{\bibfnamefont{S. P.} \bibnamefont{Domino}},
\bibnamefont{and} \bibinfo{author}{\bibfnamefont{G.}~\bibnamefont{Iaccarino}},
\bibinfo{title}{A framework for characterizing structural uncertainty in large-eddy simulation closures},
\bibinfo{journal}{Fluid, Turbul. Combust.}
\textbf{\bibinfo{volume}{100(2)}}, \bibinfo{pages}{341}
(\bibinfo{year}{2018}).

\bibitem[{\citenamefont{Ma et~al.}(2019)\citenamefont{Ma}}]{Ma2019}
\bibinfo{author}{\bibfnamefont{C.} \bibnamefont{Ma}},
\bibinfo{author}{\bibfnamefont{J.} \bibnamefont{Wang}},
\bibnamefont{and} \bibinfo{author}{\bibfnamefont{W.}~\bibnamefont{E}},
\bibinfo{title}{Model reduction with memory and the machine learning of dynamical systems},
\bibinfo{journal}{Commun. Comput. Phys.}
\textbf{\bibinfo{volume}{25(4)}}, \bibinfo{pages}{947-962}
(\bibinfo{year}{2019}).

\bibitem[{\citenamefont{Wu et~al.}(2019)\citenamefont{Wu}}]{Wu2019}
\bibinfo{author}{\bibfnamefont{J. L.} \bibnamefont{Wu}},
\bibinfo{author}{\bibfnamefont{H.} \bibnamefont{Xiao}},
\bibinfo{author}{\bibfnamefont{R.} \bibnamefont{Sun}},
\bibnamefont{and} \bibinfo{author}{\bibfnamefont{Q. Q.}~\bibnamefont{Wang}},
\bibinfo{title}{Reynolds-averaged Navier-Stokes equations with explicit data-driven Reynolds stress closure can be ill-conditioned},
\bibinfo{journal}{J. Fluid Mech.}
\textbf{\bibinfo{volume}{869}}, \bibinfo{pages}{553-586}
(\bibinfo{year}{2019}).

\bibitem[{\citenamefont{Zhu et~al.}(2019)\citenamefont{Zhu}}]{Zhu2019}
\bibinfo{author}{\bibfnamefont{L. Y.} \bibnamefont{Zhu}},
\bibinfo{author}{\bibfnamefont{W. W.} \bibnamefont{Zhang}},
\bibinfo{author}{\bibfnamefont{J. Q.} \bibnamefont{Kou}},
\bibnamefont{and} \bibinfo{author}{\bibfnamefont{Y. L.}~\bibnamefont{Liu}},
\bibinfo{title}{Machine learning methods for turbulence modeling in subsonic flows around airfoils},
\bibinfo{journal}{Phys. Fluids}
\textbf{\bibinfo{volume}{31}}, \bibinfo{pages}{015105}
(\bibinfo{year}{2010}).

\bibitem[{\citenamefont{Xie et~al.}(2019)\citenamefont{Xie}}]{Xie2019}
\bibinfo{author}{\bibfnamefont{C. Y.} \bibnamefont{Xie}},
\bibinfo{author}{\bibfnamefont{J. C.} \bibnamefont{Wang}},
\bibinfo{author}{\bibfnamefont{K.} \bibnamefont{Li}},
\bibnamefont{and} \bibinfo{author}{\bibfnamefont{C.}~\bibnamefont{Ma}},
\bibinfo{title}{Artificial neural network approach to large-eddy simulation of compressible isotropic turbulence},
\bibinfo{journal}{Phys. Rev. E}
\textbf{\bibinfo{volume}{99}}, \bibinfo{pages}{053113}
(\bibinfo{year}{2019}).

\bibitem[{\citenamefont{Xie et~al.}(2019)\citenamefont{Xie}}]{Xie2019b}
\bibinfo{author}{\bibfnamefont{C. Y.} \bibnamefont{Xie}},
\bibinfo{author}{\bibfnamefont{K.} \bibnamefont{Li}},
\bibinfo{author}{\bibfnamefont{C.}~\bibnamefont{Ma}},
\bibnamefont{and} \bibinfo{author}{\bibfnamefont{J. C.} \bibnamefont{Wang}},
\bibinfo{title}{Modeling subgrid-scale force and divergence of heat flux of compressible isotropic turbulence by artificial neural network},
\bibinfo{journal}{Phys. Rev. Fluids}
\textbf{\bibinfo{volume}{4}}, \bibinfo{pages}{104605}
(\bibinfo{year}{2019}).

\bibitem[{\citenamefont{Xie et~al.}(2019)\citenamefont{Xie}}]{Xie2019c}
\bibinfo{author}{\bibfnamefont{C. Y.} \bibnamefont{Xie}},
\bibinfo{author}{\bibfnamefont{J. C.} \bibnamefont{Wang}},
\bibinfo{author}{\bibfnamefont{H.} \bibnamefont{Li}},
\bibinfo{author}{\bibfnamefont{M. P.} \bibnamefont{Wan}},
\bibnamefont{and} \bibinfo{author}{\bibfnamefont{S. Y.}~\bibnamefont{Chen}},
\bibinfo{title}{Artificial neural network mixed model for large eddy simulation of compressible isotropic turbulence},
\bibinfo{journal}{Phys. Fluids}
\textbf{\bibinfo{volume}{31}}, \bibinfo{pages}{085112}
(\bibinfo{year}{2019}).

\bibitem[{\citenamefont{Duraisamy et~al.}(2019)\citenamefont{Duraisamy}}]{Duraisamy2019}
\bibinfo{author}{\bibfnamefont{K.} \bibnamefont{Duraisamy}},
\bibinfo{author}{\bibfnamefont{G.} \bibnamefont{Iaccarino}},
\bibnamefont{and} \bibinfo{author}{\bibfnamefont{H.}~\bibnamefont{Xiao}},
\bibinfo{title}{Turbulence modeling in the age of data},
\bibinfo{journal}{Annu. Rev. Fluid Mech.}
\textbf{\bibinfo{volume}{51}}, \bibinfo{pages}{357-377}
(\bibinfo{year}{2019}).

\bibitem[{\citenamefont{Fang et~al.}(2020)\citenamefont{Fang}}]{Fang2020}
\bibinfo{author}{\bibfnamefont{R.} \bibnamefont{Fang}},
\bibinfo{author}{\bibfnamefont{D.} \bibnamefont{Sondak}},
\bibinfo{author}{\bibfnamefont{P.} \bibnamefont{Protopapas}},
\bibnamefont{and} \bibinfo{author}{\bibfnamefont{S.}~\bibnamefont{Succi}},
\bibinfo{title}{Neural network models for the anisotropic reynolds stress tensor in turbulent channel flow},
\bibinfo{journal}{J. Turbul.}
\textbf{\bibinfo{volume}{21}}, \bibinfo{pages}{525-543}
(\bibinfo{year}{2020}).

\bibitem[{\citenamefont{Pandey et~al.}(2020)\citenamefont{Pandey}}]{Pandey2020}
\bibinfo{author}{\bibfnamefont{S.} \bibnamefont{Pandey}},
\bibinfo{author}{\bibfnamefont{J.} \bibnamefont{Schumacher}},
\bibnamefont{and} \bibinfo{author}{\bibfnamefont{K. R.}~\bibnamefont{Sreenivasan}},
\bibinfo{title}{A perspective on machine learning in turbulent flows},
\bibinfo{journal}{J. Turbul.}
\textbf{\bibinfo{volume}{21}}, \bibinfo{pages}{567-584}
(\bibinfo{year}{2020}).

\bibitem[{\citenamefont{Duraisamy et~al.}(2020)\citenamefont{Duraisamy}}]{Duraisamy2020}
\bibinfo{author}{\bibfnamefont{K.} \bibnamefont{Duraisamy}},
\bibinfo{title}{Machine Learning-augmented Reynolds-averaged $\&$ Large Eddy Simulation Models of Turbulence},
\bibinfo{journal}{arXiv:2009.10675}
(\bibinfo{year}{2020}).

\bibitem[{\citenamefont{Pawar et~al.}(2020)\citenamefont{Pawar}}]{Pawar2020}
\bibinfo{author}{\bibfnamefont{S.} \bibnamefont{Pawar}},
\bibinfo{author}{\bibfnamefont{O.} \bibnamefont{San}},
\bibinfo{author}{\bibfnamefont{B.} \bibnamefont{Aksoylu}},
\bibinfo{author}{\bibfnamefont{A.} \bibnamefont{Rasheed}},
\bibnamefont{and} \bibinfo{author}{\bibfnamefont{T.}~\bibnamefont{Kvamsal}},
\bibinfo{title}{Physcis guided machine learning using simplified theories},
\bibinfo{journal}{Arxiv:2012:13343v1}
(\bibinfo{year}{2020}).

\bibitem[{\citenamefont{Beetham et~al.}(2020)\citenamefont{Beetham}}]{Beetham2020}
\bibinfo{author}{\bibfnamefont{S.} \bibnamefont{Beetham}},
\bibnamefont{and} \bibinfo{author}{\bibfnamefont{J.}~\bibnamefont{JCapecelatro}},
\bibinfo{title}{Formulating turbulence closures using sparse regression with embedded form invariance},
\bibinfo{journal}{Phys. Rev. Fluids}
\textbf{\bibinfo{volume}{5}}, \bibinfo{pages}{084611}
(\bibinfo{year}{2020}).

\bibitem[{\citenamefont{Samtaney et~al.}(2001)\citenamefont{Samtany}}]{Samtaney2001}
\bibinfo{author}{\bibfnamefont{R.} \bibnamefont{Samtaney}},
\bibinfo{author}{\bibfnamefont{D. I.} \bibnamefont{Pullin}},
\bibnamefont{and} \bibinfo{author}{\bibfnamefont{B.}~\bibnamefont{Kosovi$\acute{c}$}},
\bibinfo{title}{Direct numerical simulation of decaying compressible turbulence and shocklet statistics},
\bibinfo{journal}{Phys. Fluids}
\textbf{\bibinfo{volume}{13}}, \bibinfo{pages}{1415}
(\bibinfo{year}{2001}).

\bibitem[{\citenamefont{Li et~al.}(2010)\citenamefont{Li}}]{Li2010}
\bibinfo{author}{\bibfnamefont{X. L.} \bibnamefont{Li}},
\bibinfo{author}{\bibfnamefont{D. X.} \bibnamefont{Fu}},
\bibnamefont{and} \bibinfo{author}{\bibfnamefont{Y. W.}~\bibnamefont{Ma}},
\bibinfo{title}{Direct numerical simulation of compressible turbulent flows},
\bibinfo{journal}{Acta. Mech. Sin.}
\textbf{\bibinfo{volume}{26}}, \bibinfo{pages}{795-806}
(\bibinfo{year}{2010}).

\bibitem[{\citenamefont{Wang et~al.}(2012)\citenamefont{Wang}}]{Wang2012}
\bibinfo{author}{\bibfnamefont{J.} \bibnamefont{Wang}},
\bibinfo{author}{\bibfnamefont{Y.} \bibnamefont{Shi}},
\bibinfo{author}{\bibfnamefont{L.-P.} \bibnamefont{Wang}},
\bibinfo{author}{\bibfnamefont{Z.} \bibnamefont{Xiao}},
\bibinfo{author}{\bibfnamefont{X. T.} \bibnamefont{He}},
\bibnamefont{and} \bibinfo{author}{\bibfnamefont{S.}~\bibnamefont{Chen}},
\bibinfo{title}{Effect of compressibility on the small scale structures in isotropic turbulence},
\bibinfo{journal}{J. Fluid Mech.}
\textbf{\bibinfo{volume}{713}}, \bibinfo{pages}{588}
(\bibinfo{year}{2012}).

\bibitem[{\citenamefont{Balakumar et~al.}(2015)\citenamefont{Balakumar}}]{Balakumar2015}
\bibinfo{author}{\bibfnamefont{P.} \bibnamefont{Balakumar}},
\bibinfo{title}{DNS/LES simulations of separated flows at high Reynolds numbers},
\bibinfo{journal}{45th AIAA Fluid Dynamics Conference, (Dallas, Texas)}
\textbf{\bibinfo{volume}{}}, \bibinfo{pages}{2015-2783}
(\bibinfo{year}{2015}).

\bibitem[{\citenamefont{Gloerfelt et~al.}(2015)\citenamefont{Gloerfelt}}]{Gloerfelt2015}
\bibinfo{author}{\bibfnamefont{X.} \bibnamefont{Gloerfelt}},
\bibnamefont{and} \bibinfo{author}{\bibfnamefont{P.}~\bibnamefont{Cinnella}},
\bibinfo{title}{Investigation of the flow dynamics in a channel constricted by periodic hills},
\bibinfo{journal}{45th AIAA Fluid Dynamics Conference, (Dalla, Texas)}
\textbf{\bibinfo{volume}{}}, \bibinfo{pages}{2015-2480}
(\bibinfo{year}{2015}).

\bibitem[{\citenamefont{Sutherland et~al.}(1992)\citenamefont{Sutherland}}]{Sutherland1992}
\bibinfo{author}{\bibfnamefont{W.} \bibnamefont{Sutherland}},
\bibinfo{title}{The viscosity of gases and molecular force},
\bibinfo{journal}{Philos. Mag. S5 }
\textbf{\bibinfo{volume}{36}}, \bibinfo{pages}{507-531}
(\bibinfo{year}{1992}).

\bibitem[{\citenamefont{Favre et~al.}(1965)\citenamefont{Favre}}]{Favre1965}
\bibinfo{author}{\bibfnamefont{A.} \bibnamefont{Favre}},
\bibinfo{title}{Equations des gaz turbulents compressible. {I}. {F}ormes generales},
\bibinfo{journal}{J. Mec.}
\textbf{\bibinfo{volume}{4}}, \bibinfo{pages}{361}
(\bibinfo{year}{1965}).

\bibitem[{\citenamefont{Jiang et~al.}(2013)\citenamefont{Jiang}}]{Jiang2013}
\bibinfo{author}{\bibfnamefont{Z.} \bibnamefont{Jiang}},
\bibinfo{author}{\bibfnamefont{Z. L.} \bibnamefont{Xiao}},
\bibinfo{author}{\bibfnamefont{Y. P.} \bibnamefont{Shi}},
\bibnamefont{and} \bibinfo{author}{\bibfnamefont{S. Y.}~\bibnamefont{Chen}},
\bibinfo{title}{Constrained large-eddy simulation of wall-bounded compressible turbulent flows},
\bibinfo{journal}{Phys. Fluids}
\textbf{\bibinfo{volume}{25}}, \bibinfo{pages}{106102}
(\bibinfo{year}{2013}).

\bibitem[{\citenamefont{Xia et~al.}(2013)\citenamefont{Xia}}]{Xia2013}
\bibinfo{author}{\bibfnamefont{Z. H.} \bibnamefont{Xia}},
\bibinfo{author}{\bibfnamefont{Y. P.} \bibnamefont{Shi}},
\bibinfo{author}{\bibfnamefont{R. K.} \bibnamefont{Hong}},
\bibinfo{author}{\bibfnamefont{Z. L.} \bibnamefont{Xiao}},
\bibnamefont{and} \bibinfo{author}{\bibfnamefont{S. Y.}~\bibnamefont{Chen}},
\bibinfo{title}{Constrained large-eddy simulation of separated flow in a channel with streamwise-periodic constrictions},
\bibinfo{journal}{J. Turb.}
\textbf{\bibinfo{volume}{14}}, \bibinfo{pages}{1-21}
(\bibinfo{year}{2013}).

\bibitem[{\citenamefont{Mellen et~al.}(2000)\citenamefont{Mellen}}]{Mellen2000}
\bibinfo{author}{\bibfnamefont{C. P.} \bibnamefont{Mellen}},
\bibinfo{author}{\bibfnamefont{J.} \bibnamefont{Fr$\ddot{o}$hlich}},
\bibnamefont{and} \bibinfo{author}{\bibfnamefont{W.}~\bibnamefont{Rodi}},
\bibinfo{title}{Large-eddy simulation of the flow over periodic hills,In: Deville M, Owens R, editors},
\bibinfo{journal}{Proceedings of 16th IMACS world congress, Lausanne, Switzerland. CD-ROM}
(\bibinfo{year}{2000}).

\bibitem[{\citenamefont{Frohlich et~al.}(2005)\citenamefont{Frohlich}}]{Frohlich2005}
\bibinfo{author}{\bibfnamefont{J.} \bibnamefont{Frohlich}},
\bibinfo{author}{\bibfnamefont{C. P.} \bibnamefont{Mellen}},
\bibinfo{author}{\bibfnamefont{W.} \bibnamefont{Rodi}},
\bibinfo{author}{\bibfnamefont{L.} \bibnamefont{Temmerman}},
\bibnamefont{and} \bibinfo{author}{\bibfnamefont{M. A.}~\bibnamefont{Leschziner}},
\bibinfo{title}{Highly Resolved Large-Eddy Simulation of Separated Flow in a Channel with Streamwise Periodic Constraint},
\bibinfo{journal}{J. Fluid Mech.}
\textbf{\bibinfo{volume}{526}}, \bibinfo{pages}{19-66}
(\bibinfo{year}{2005}).

\bibitem[{\citenamefont{Fu et~al.}(2010)\citenamefont{Fu}}]{Fu2010}
\bibinfo{author}{\bibfnamefont{D. X.}~\bibnamefont{Fu}},
\bibinfo{author}{\bibfnamefont{Y. W.} \bibnamefont{Ma}},
\bibinfo{author}{\bibfnamefont{X. L.} \bibnamefont{Li}},
\bibnamefont{and} \bibinfo{author}{\bibfnamefont{Q.}~\bibnamefont{Wang}},
\bibinfo{book}{{\it Direct Numerical Simulation of Compressible Turbulence}}
(\bibinfo{year}{Science Press, Beijing (in Chinese), 2010}).

\bibitem[{\citenamefont{Lele et~al.}(1992)\citenamefont{Lele}}]{Lele1992}
\bibinfo{author}{\bibfnamefont{S. K.} \bibnamefont{Lele}},
\bibinfo{title}{Compact finite difference schemes with spectral-like resolution},
\bibinfo{journal}{J. Comput. Phys.}
\textbf{\bibinfo{volume}{103}}, \bibinfo{pages}{16}
(\bibinfo{year}{1992}).

\bibitem[{\citenamefont{Obayashi et~al.}(1986)\citenamefont{Obayashi}}]{Obayashi1986}
\bibinfo{author}{\bibfnamefont{S.} \bibnamefont{Obayashi}},
\bibinfo{author}{\bibfnamefont{K.} \bibnamefont{Matsushima}},
\bibinfo{author}{\bibfnamefont{K.} \bibnamefont{Fujii}},
\bibnamefont{and} \bibinfo{author}{\bibfnamefont{K.}~\bibnamefont{Kuwahara}},
\bibinfo{title}{Improvements in efficiency and reliability for Navier-Stokes compuataions using the LU-ADI factorization algorithm},
\bibinfo{journal}{aiaa}
\textbf{\bibinfo{volume}{0338}}, \bibinfo{pages}{}
(\bibinfo{year}{1986}).

\bibitem[{\citenamefont{Ziefle et~al.}(2008)\citenamefont{Ziefle}}]{Ziefle2008}
\bibinfo{author}{\bibfnamefont{J.} \bibnamefont{Ziefle}},
\bibinfo{author}{\bibfnamefont{S.} \bibnamefont{Stolz}},
\bibnamefont{and} \bibinfo{author}{\bibfnamefont{L.}~\bibnamefont{Kleiser}},
\bibinfo{title}{Large-Eddy Simulation of Separated Flow in a Channel with StreamwisePeriodic Constrictions},
\bibinfo{journal}{AIAA Journal}
\textbf{\bibinfo{volume}{46(7)}}, \bibinfo{pages}{1705-1718}
(\bibinfo{year}{2008}).

\bibitem[{\citenamefont{Breuer et~al.}(2009)\citenamefont{Breuer}}]{Breuer2009}
\bibinfo{author}{\bibfnamefont{M.} \bibnamefont{Breuer}},
\bibinfo{author}{\bibfnamefont{N.} \bibnamefont{Peller}},
\bibinfo{author}{\bibfnamefont{Ch.} \bibnamefont{Rapp}},
\bibnamefont{and} \bibinfo{author}{\bibfnamefont{M.}~\bibnamefont{Manhart}},
\bibinfo{title}{Flow over periodic hills - Numerical and experimental study in a wide range of Reynolds numbers},
\bibinfo{journal}{Computer $\&$ Fluids}
\textbf{\bibinfo{volume}{38}}, \bibinfo{pages}{433-457}
(\bibinfo{year}{2009}).

\bibitem[{\citenamefont{Ge et~al.}(2019)\citenamefont{Ge}}]{Ge2019}
\bibinfo{author}{\bibfnamefont{X.} \bibnamefont{Ge}},
\bibinfo{author}{\bibfnamefont{O. V.} \bibnamefont{Vasilyev}},
\bibnamefont{and} \bibinfo{author}{\bibfnamefont{M. Y.}~\bibnamefont{Hussaini}},
\bibinfo{title}{Wavelet-based adaptive delayed detached eddy simulations for wall-bounded compressible turbulent flows},
\bibinfo{journal}{J. Fluid Mech.}
\textbf{\bibinfo{volume}{873}}, \bibinfo{pages}{1116-1157}
(\bibinfo{year}{2019}).

\bibitem[{\citenamefont{Gamahara et~al.}(2017)\citenamefont{Gamahara}}]{Gamahara2017}
\bibinfo{author}{\bibfnamefont{M.} \bibnamefont{Gamahara}},
\bibnamefont{and} \bibinfo{author}{\bibfnamefont{Y.}~\bibnamefont{Hattori}},
\bibinfo{title}{Searching for turbulence models by artificial neural network},
\bibinfo{journal}{Phys. Rev. Fluids}
\textbf{\bibinfo{volume}{2(5)}}, \bibinfo{pages}{054604}
(\bibinfo{year}{2017}).

\bibitem[{\citenamefont{Maulik et~al.}(2020)\citenamefont{Maulik}}]{Maulik2020}
\bibinfo{author}{\bibfnamefont{R.} \bibnamefont{Maulik}},
\bibinfo{author}{\bibfnamefont{H.} \bibnamefont{Sharma}},
\bibinfo{author}{\bibfnamefont{S.} \bibnamefont{Patel}},
\bibinfo{author}{\bibfnamefont{B.} \bibnamefont{Lusch}},
\bibnamefont{and} \bibinfo{author}{\bibfnamefont{E.}~\bibnamefont{Jennings}},
\bibinfo{title}{Accelerating RANS simulations using a datadriven framework for eddy-viscosity emulation},
\bibinfo{journal}{arxiv:1910.10878v4}
(\bibinfo{year}{2020}).

\bibitem[{\citenamefont{Xie et~al.}(2020)\citenamefont{Xie}}]{Xie2020}
\bibinfo{author}{\bibfnamefont{C. Y.} \bibnamefont{Xie}},
\bibinfo{author}{\bibfnamefont{J. C.} \bibnamefont{Wang}},
\bibnamefont{and} \bibinfo{author}{\bibfnamefont{W. N.}~\bibnamefont{E}},
\bibinfo{title}{Modeling subgrid-scale forces by spatial artificial neural networks in large eddy simulation of turbulence},
\bibinfo{journal}{Phys. Rev. Fluids}
\textbf{\bibinfo{volume}{5}}, \bibinfo{pages}{054606}
(\bibinfo{year}{2020}).

\bibitem[{\citenamefont{Zhang et~al.}(1998)\citenamefont{Zhang}}]{Zhang1998}
\bibinfo{author}{\bibfnamefont{G.} \bibnamefont{Zhang}},
\bibinfo{author}{\bibfnamefont{B. E.} \bibnamefont{Patuwo}},
\bibnamefont{and} \bibinfo{author}{\bibfnamefont{M. Y.}~\bibnamefont{Hu}},
\bibinfo{title}{Forecasting with artificial neural networks: the state of the art},
\bibinfo{journal}{Intl J. Forecast.}
\textbf{\bibinfo{volume}{14}}, \bibinfo{pages}{35}
(\bibinfo{year}{1998}).

\bibitem[{\citenamefont{Demuth et~al.}(2014)\citenamefont{Demuth}}]{Demuth2014}
\bibinfo{author}{\bibfnamefont{H. B.} \bibnamefont{Demuth}},
\bibinfo{author}{\bibfnamefont{M. H.} \bibnamefont{Beale}},
\bibinfo{author}{\bibfnamefont{O. D.} \bibnamefont{Jess}},
\bibnamefont{and} \bibinfo{author}{\bibfnamefont{M. T.}~\bibnamefont{Hagan}},
\bibinfo{title}{Neural Network Design},
\bibinfo{journal}{Stillwater, OK: Martin Hagan}
(\bibinfo{year}{2014}).

\bibitem[{\citenamefont{Maulik et~al.}(2017)\citenamefont{Maulik}}]{Maulik2017}
\bibinfo{author}{\bibfnamefont{R.} \bibnamefont{Maulik}},
\bibnamefont{and} \bibinfo{author}{\bibfnamefont{O.}~\bibnamefont{San}},
\bibinfo{title}{A neural network approach for the blind deconvolution of turbulent flows},
\bibinfo{journal}{J. Fluid Mech.}
\textbf{\bibinfo{volume}{831}}, \bibinfo{pages}{151}
(\bibinfo{year}{2017}).

\bibitem[{\citenamefont{Kingma et~al.}(2014)\citenamefont{Kingma}}]{Kingma2014}
\bibinfo{author}{\bibfnamefont{D. P.} \bibnamefont{Kingma}},
\bibnamefont{and} \bibinfo{author}{\bibfnamefont{J.}~\bibnamefont{Ba}},
\bibinfo{title}{Adam: A method for stochastic optimization},
\bibinfo{journal}{arXiv:1412.6980}
(\bibinfo{year}{2014}).

\bibitem[{\citenamefont{Xiao et~al.}(2020)\citenamefont{Xiao}}]{Xiao2020}
\bibinfo{author}{\bibfnamefont{H.} \bibnamefont{Xiao}},
\bibinfo{author}{\bibfnamefont{J. L.} \bibnamefont{Wu}},
\bibinfo{author}{\bibfnamefont{S.} \bibnamefont{Laizet}},
\bibnamefont{and} \bibinfo{author}{\bibfnamefont{L.}~\bibnamefont{Duan}},
\bibinfo{title}{Flows over periodic hills of parameterized geometries: A dataset for data-driven turbulence modeling from direct simulations},
\bibinfo{journal}{Computers $\&$ Fluids}
\textbf{\bibinfo{volume}{200}}, \bibinfo{pages}{104431}
(\bibinfo{year}{2020}).

\bibitem[{\citenamefont{Gao et~al.}(2005)\citenamefont{Gao}}]{Gao2005}
\bibinfo{author}{\bibfnamefont{H.} \bibnamefont{Gao}},
\bibinfo{author}{\bibfnamefont{D. X.} \bibnamefont{Fu}},
\bibinfo{author}{\bibfnamefont{Y. W.} \bibnamefont{Ma}},
\bibnamefont{and} \bibinfo{author}{\bibfnamefont{X. L.}~\bibnamefont{Li}},
\bibinfo{title}{Direct numerical simulation of supersonic boundary layer flow},
\bibinfo{journal}{Chinese Phys. Lett.}
\textbf{\bibinfo{volume}{22}}, \bibinfo{pages}{1709-1712}
(\bibinfo{year}{2005}).

\bibitem[{\citenamefont{Li et~al.}(2005)\citenamefont{Gao}}]{Li2005}
\bibinfo{author}{\bibfnamefont{X. L.} \bibnamefont{Li}},
\bibinfo{author}{\bibfnamefont{D. X.} \bibnamefont{Fu}},
\bibnamefont{and} \bibinfo{author}{\bibfnamefont{Y. W.}~\bibnamefont{Ma}},
\bibinfo{title}{DNS of compressible turbulent boundary layer over a blunt wedge},
\bibinfo{journal}{Sci. China Ser. G.}
\textbf{\bibinfo{volume}{48}}, \bibinfo{pages}{129-141-1712}
(\bibinfo{year}{2005}).

\bibitem[{\citenamefont{Rapp et~al.}(2011)\citenamefont{Rapp}}]{Rapp2011}
\bibinfo{author}{\bibfnamefont{C.} \bibnamefont{Rapp}},
\bibnamefont{and} \bibinfo{author}{\bibfnamefont{M.}~\bibnamefont{Manhart}},
\bibinfo{title}{Flow over periodic hills: an experimental study},
\bibinfo{journal}{Exp. Fluids}
\textbf{\bibinfo{volume}{51}}, \bibinfo{pages}{247-269}
(\bibinfo{year}{2011}).

\end{thebibliography}

\end{document}